\documentclass{aa}  
\usepackage{graphicx}
\usepackage{natbib}
\usepackage{txfonts}
\begin{document}
\title{An interferometric study of the low-mass protostar
  IRAS~16293-2422: small scale organic chemistry}

\author{S.~E. Bisschop\inst{1}\thanks{Current address:
    Max-Planck-Institut f{\"u}r Radioastronomie, Auf dem H{\"u}gel 69,
    53121 Bonn, Germany} \and J.~K. J{\o}rgensen\inst{2} \and
  T.~L. Bourke\inst{3} \and S. Bottinelli\inst{1} \and E.~F. van
  Dishoeck\inst{1}$^,$\inst{4}}

\offprints{S.~E. Bisschop, bisschop@mpifr-bonn.mpg.de}

\institute{Leiden Observatory, Leiden University, P.~O. Box 9513, 2300
  RA Leiden, Netherlands \and {Argelander-Institut f{\"u}r Astronomie,
    University of Bonn, Auf dem H{\"u}gel 71, 53121 Bonn, Germany}
  \and {Harvard-Smithsonian Center for Astrophysics, 60 Garden Street,
    Cambridge, MA 02138, USA} \and {Max-Planck-Institut f{\"u}r
    Extraterrestische Physik, Giessenbachstrasse 1, 85748 Garching,
    Germany}}

   \date{Received; accepted}

 
  \abstract
  {}
  {To investigate the chemical relations between complex organics
    based on their spatial distributions and excitation conditions in
    the low-mass young stellar objects IRAS~16293-2422 ``A'' and
    ``B''.}
  {Interferometric observations with the Submillimeter Array have been
    performed at 5$\arcsec\times3\arcsec$\ (800$\times$500~AU)
    resolution revealing emission lines of HNCO, CH$_3$CN, CH$_2$CO,
    CH$_3$CHO and C$_2$H$_5$OH. Rotational temperatures are determined
    from rotational diagrams when a sufficient number of lines are
    detected.}
  {Compact emission is detected for all species studied here. For HNCO
    and CH$_3$CN it mostly arises from source ``A'', CH$_2$CO and
    C$_2$H$_5$OH have comparable strength for both sources and
    CH$_3$CHO arises exclusively from source ``B''. HNCO, CH$_3$CN and
    CH$_3$CHO have rotational temperatures $>$200~K implying that they
    arise from hot gas. The $(u,v)$-visibility data reveal that HNCO
    also has extended cold emission, which could not be previously
    determined through single dish data. }
  {The relative abundances of the molecules studied here are very
    similar within factors of a few to those found in high-mass
    YSOs. This illustrates that the chemistry between high- and
    low-mass objects appears to be relatively similar and thus
    independent of luminosity and cloud mass. In contrast, bigger
    abundance differences are seen between the ``A'' and ``B''
    source. For instance, the HNCO abundance relative to CH$_3$OH is
    $\sim$4 times higher toward ``A'', which may be due to a higher
    initial OCN$^-$ ice abundances in source ``A'' compared to
    ``B''. Furthermore, not all oxygen-bearing species are
    co-existent, with CH$_3$CHO/CH$_3$OH an order of magnitude higher
    toward ``B'' than ``A''. The different spatial behavior of
    CH$_2$CO and C$_2$H$_5$OH compared with CH$_3$CHO suggests that
    successive hydrogenation reactions on grain-surfaces are not
    sufficient to explain the observed gas phase abundance of the
    latter. Selective destruction of CH$_3$CHO may result in the
    anti-coincidence of these species in source ``A''. These results
    illustrate the power of interferometric compared with single dish
    data in terms of testing chemical models.}

  \keywords{Astrochemistry - Line:identification -
    Methods:observational - Techniques:interferometric -
    Stars:formation}

  \authorrunning{S.~E. Bisschop et al.}  \titlerunning{Small scale
    organic chemistry in IRAS~16293-2422}
   \maketitle

\section{Introduction}

The envelopes of some low-mass protostars contain many complex organic
molecules\footnote{In this paper molecules are considered complex if
  they contain more than four atoms.}
\citep{blake1994,dishoeck1995,cazaux2003,bottinelli2004b,bottinelli2004,bottinelli2007,jorgensen2005b,jorgensen2005a,sakai2006,sakai2007}.
This raises the question whether these are low-mass versions of ``hot
cores'', chemically very rich environments in high-mass star forming
regions that are thought to have their origin in grain-mantle
evaporation and the subsequent rapid gas phase reactions. The low-mass
counterpart is sometimes called a ``hot corino''. The presence of warm
material has long been suggested from modeling of the SEDs of these
sources \citep[e.g.,][]{adams1987,jorgensen2002,shirley2002} and has
been firmly established by molecular excitation studies
\citep{blake1994,dishoeck1995,ceccarelli2000} and by resolved
interferometric imaging \citep[see
e.g.,][]{chandler2005,bottinelli2004,jorgensen2005}. However, in some
sources the emission peaks offset from the continuum source
\citep[e.g.,][]{chandler2005}. This offset is in disagreement with the
``hot corino'' hypothesis for low-mass stars in which complex
molecules evaporate through passive heating and is more in favor of
other explanations such as the presence of disks or outflows that
create shocks in the envelope. Currently, there is an ongoing debate
on whether the emission of complex organics comes from ``hot corinos''
or from other types of regions. Also, the extent to which the observed
organics are first generation molecules created in the ice or second
generation produced in the gas is still an open question. The aim of
this work is to map the emission of complex organics to address the
latter question, namely to determine their most likely formation
mechanism.

One method to study chemical links between species is to study
molecular abundances through single-dish surveys in a large number of
sources \citep{vdtak2000b,vdtak2003,ikeda2001,bisschop2007a}. An
alternative method is to look for spatial correlations by
interferometric observations of a single source through which it is
possible to distinguish compact and extended emission as well as the
exact location of the compact emission. The species that are studied
here are the nitrogen-bearing species, HNCO and CH$_3$CN and the
oxygen-bearing species CH$_2$CO, CH$_3$CHO, and C$_2$H$_5$OH that are
commonly found toward or surrounding hot cores. The molecules
CH$_2$CO, CH$_3$CHO and C$_2$H$_5$OH are proposed to be linked by
successive hydrogenation on the surfaces of grains
\citep{tielens1997}.

The source studied in this paper, IRAS~16293-2422, is a very well
studied and chemically rich low-mass YSO. It is a binary with its main
components, named ``A'' and ``B'', separated by 5\arcsec
\citep[corresponding to 800~AU at 160~pc][]{mundy1992}. From single
dish observations and modeling, \citet{dishoeck1995},
\citet{ceccarelli1999,ceccarelli2000} and \citet{schoier2002}
concluded that emission from several organic molecules arises from a
compact region. The data for some species such as H$_2$CO and CH$_3$CN
are significantly better fitted if a ``jump'' in the abundance at
80--90~K due to grain-mantle evaporation is assumed in a spherical
circum-binary envelope \citep{ceccarelli2000,schoier2002}. Through
interferometric observations it is possible to distinguish between the
emission from the cold extended envelope and more compact emission as
well as the peak location: is the emission coming from source ``A'' or
``B''?  Previous observations have shown that some complex species are
much more abundant toward one source than the other, such as
CH$_3$OCHO which is more prominent toward the ``A'' source
\citep[][]{bottinelli2004,huang2005,kuan2004,remijan2006}. 

The nitrogen-bearing species HNCO and CH$_3$CN have previously been
detected toward IRAS~16293-2422 through single dish observations
\citep{dishoeck1995,cazaux2003}. They often have high rotational
temperatures in star forming regions implying that they are present in
hot gas \citep{olmi1993,zinchenko2000,cazaux2003,bisschop2007a}. This
is also the case for the oxygen-bearing molecule C$_2$H$_5$OH
\citep{ikeda2001,bisschop2007a}. In contrast, CH$_2$CO and CH$_3$CHO
are often detected in high-mass sources with low rotational
temperatures \citep{ikeda2002,bisschop2007a}. For CH$_3$CHO higher
rotational temperatures are also found, but these can be due to the
{\it b}-type transitions that are radiatively pumped and do not
represent the actual kinetic temperature of the gas
\citep{nummelin2000,turner1991}. Recent laboratory experiments by
\citet{bisschop2007c} have shown that it is possible to explain the
gas phase abundances of C$_2$H$_5$OH if a solid state formation route
through CH$_3$CHO hydrogenation is assumed. The observed absence of
CH$_3$CHO in hot gas combined with its detection in cold ices
\citep{keane2001,gibb2004}, however, implies that it must be destroyed
at higher ice temperatures, before evaporation commences, or directly
after evaporation in the gas phase. A comparison of the excitation
properties as well as interferometric observations in which the
spatial distribution of the species can be determined are good tools
to further elucidate the chemical relations between these species.

This paper is structured as follows: \S~\ref{obs} presents the
Submillimeter Array (SMA) observing strategy as well as maps and
spectra; \S~\ref{an} explains the analysis of the interferometric
observations in the $(u,v)$-plane and the fitting of rotational
diagrams; \S~\ref{results} presents the results of the rotational
diagram and flux analysis; \S~\ref{disc} discusses the relative
abundances of the different species with respect to each other as well
as the implications for the chemistry; \S~\ref{sum} summarizes the
main conclusions.

\section{Observations}
\label{obs}
\subsection{Observing strategy}
The line-rich low-mass protostar, IRAS~16293-2422, at a distance of
160~pc, has been surveyed from 2004 to 2005 using the Submillimeter
Array\footnote{The Submillimeter Array is a joint project between the
  Smithsonian Astrophysical Observatory and the Academia Sinica
  Institute of Astronomy and Astrophysics. It is funded by the
  Smithsonian Institute and Academia Sinica.} (SMA) in a large number
of frequency settings. In this paper we focus on just the emission
lines from HNCO, CH$_3$CN, C$_2$H$_5$OH, CH$_2$CO and CH$_3$CHO for
which particularly sensitive data exist (see Tables~\ref{sma_flux} and
\ref{sma_flux2}). Other papers presenting data from this survey are
\cite{takakuwa2007} and \cite{yeh2008}.

The data shown here were all covered in one spectral setup at
219.4--221.3~GHz (LSB) and 229.4--231.3~GHz (USB) on February 18th
2005 with 6 antennas in the compact configuration of the array,
resulting in 15 baselines from 10--70~m. This frequency setup was
chosen to complement a large SMA survey of Class 0 sources
\citep{jorgensen2007} and includes, e.g., the strong CO transitions
and its isotopologues and a large number of lines from organic
molecules. The data were taken as part of dual receiver observations
simultaneously with high frequency 690~GHz data (T.~Bourke et al., in
prep.). The weather was optimal for the 690~GHz observations with
$\tau_{\rm 225~GHz}$ better than 0.03. The excellent weather
conditions and the dual receiver option naturally provides a
possibility for very high quality observations at the lower
frequencies. The overall flux calibration is estimated to be 30\%. The
phase center for the observations, $(\alpha,\delta)_{\rm
  J2000}$=(16:32:22.9,~-24:28:35.5), is located about 1$\arcsec$ to
the north from the ``A'' component of the IRAS~16293-2422 binary. The
synthesized beam was $5.5''\times 3.2''$, which corresponds to
800$\times$500~AU at 160~pc. The SMA correlator was set up with
uniform coverage of the 2~GHz bandwidth with 128 channels for each of
the 24 chunks, corresponding to a spectral resolution of
1.1~km~s$^{-1}$. The RMS was 60~mJy~beam$^{-1}$~channel$^{-1}$.

The initial data reduction was performed using the MIR package
\citep{qi2006}. The complex gains were calibrated by frequent
observations of the quasar J1743-038 (3.0~Jy) offset by 27~degrees
from IRAS~16293-2422. Flux and band-pass calibration was performed by
observations of Uranus. Continuum subtracted line maps and further
analysis were subsequently made using the Miriad package.

\begin{table*}
  \caption{Line parameters and fluxes from detected HNCO and CH$_3$CN transitions for the ``A'' and ``B'' components. The figures in which the lines are shown are indicated in the last column.}\label{sma_flux}
\begin{center}
\begin{tabular}{llllllll}
\hline
\hline
Molecule & Freq.  & Transition & $E_{\rm u}$ & $\mu^2S$ & \multicolumn{2}{c}{\underline{$F\Delta V^a$ (Jy~km~s$^{-1}$) [$\sigma$]}} & Figure\\
         & (GHz)  &            & (K)        & (D$^2$) &  \phantom{1}``A''&  \phantom{1}``B'' \\
\hline
HNCO     & 219.657 & 10$_{3,7/8}$--9$_{3,6/7}$ & 447 &  \phantom{1}139.2 & \phantom{1}4.9\phantom{$^b$} [22$\sigma$] & \phantom{$>$}0.54 [\phantom{1}8$\sigma$] & \ref{ap1}\\
         & 219.733 & 10$_{2,9}$--9$_{2,8}$    & 231 & \phantom{11}73.8 & 10.5$^b$ [41$\sigma$] & \phantom{$>$}0.37 [\phantom{1}6$\sigma$] & \ref{ap1}\\
         & 219.737 & 10$_{2,8}$--9$_{2,7}$    & 231 & \phantom{11}73.8 & 10.5$^b$ [41$\sigma$] & \phantom{$>$}0.30 [\phantom{1}5$\sigma$] & \ref{ap1}\\
         & 219.798 & 10$_{0,10}$--9$_{0,9}$   & \phantom{2}58 & \phantom{11}77.0 & 10.9\phantom{$^b$} [49$\sigma$] & \phantom{$>$}1.37 [12$\sigma$] & \ref{overview}\\
         & 220.584 & 10$_{1,9}$--9$_{1,8}$    & 102 & \phantom{11}76.2 & \phantom{1}9.6\phantom{$^b$} [43$\sigma$] & \phantom{$>$}1.19 [13$\sigma$] & \ref{ch3c13n_spec}\\
\hline
CH$_3$CN & 220.476 & 12$_8$--11$_8$ & 526 & \phantom{1}608.6 & \phantom{1}2.4\phantom{$^b$} [11$\sigma$] & \phantom{$>$}0.45 [\phantom{1}7$\sigma$] & \ref{ap1}\\
         & 220.539 & 12$_7$--11$_7$ & 419 & \phantom{1}722.8 & \phantom{1}3.1\phantom{$^b$} [14$\sigma$] & \phantom{$>$}0.44 [\phantom{1}7$\sigma$] & \ref{ap1}\\
         & 220.594 & 12$_6$--11$_6$ & 326 & 1643.3 & \phantom{1}--$^c$ & \phantom{$>$}1.56 [14$\sigma$] & \ref{ch3c13n_spec}\\
         & 220.641 & 12$_5$--11$_5$ & 247 & \phantom{1}905.3 & \phantom{1}--$^d$ & \phantom{$>$}1.81 [19$\sigma$] & \ref{ch3c13n_spec}\\
         & 220.679 & 12$_4$--11$_4$ & 183 & \phantom{1}973.9 & 11.4\phantom{$^b$} [52$\sigma$] & \phantom{$>$}1.76 [19$\sigma$] & \ref{ap2}\\
         & 220.709 & 12$_3$--11$_3$ & 133 & 2054.4 & 14.8\phantom{$^b$} [67$\sigma$] & \phantom{$>$}2.21 [23$\sigma$] & \ref{ap2}\\
         & 220.730 & 12$_2$--11$_2$ & \phantom{1}97 & 1065.1 & 14.3\phantom{$^b$} [65$\sigma$] & \phantom{$>$}2.07 [22$\sigma$] & \ref{ap2}\\ 
         & 220.742 & 12$_1$--11$_1$ & \phantom{1}76 & 1088.0 & \phantom{1}--$^d$ & \phantom{$>$}2.23 [24$\sigma$] & \ref{ap2}\\
         & 220.747 & 12$_0$--11$_0$ & \phantom{1}69 & 1095.6 & \phantom{1}--$^d$ & \phantom{$>$}2.47 [26$\sigma$] & \ref{overview}, \ref{ap2}\\
\hline 
CH$_3^{13}$CN & 220.600 & 12$_3$--11$_3$ & 133 & 2066.3 & \phantom{1}--$^d$ & \phantom{$>$}0.88 [\phantom{1}9$\sigma$] & \ref{ch3c13n_spec}\\
             & 220.621 & 12$_2$--11$_2$ & \phantom{1}97 & 1071.4 & \phantom{1}--$^d$ & \phantom{$>$}0.69 [\phantom{1}7$\sigma$] & \ref{ch3c13n_spec}\\
             & 220.634 & 12$_1$--11$_1$ & \phantom{1}76 & 1094.4 & \phantom{1}--$^d$ & \phantom{$>$}0.61 [\phantom{1}6$\sigma$] & \ref{ch3c13n_spec}\\
             & 220.638 & 12$_0$--11$_0$ & \phantom{1}69 & 1102.1 & \phantom{1}--$^d$ & \phantom{$>$}0.44 [\phantom{1}5$\sigma$] & \ref{ch3c13n_spec}\\
\hline
\end{tabular}
\end{center}

$^aF\Delta V$ is determined over the full width of the emission coming from source ``A'' or ``B'', respectively. $^b$The 10$_{2,9}$--9$_{2,8}$ and 10$_{2,8}$--9$_{2,7}$ transitions for HNCO are blended in source ``A''. The flux given is the total combined flux for both lines. $^c$Blended with the C$_2$H$_5$OH 13$_{1,13}$--12$_{0,12}$ transition. $^d$Strongly blended lines. 
\end{table*}

\begin{table*}
  \caption{Line parameters and fluxes from detected CH$_2$CO, CH$_3$CHO and C$_2$H$_5$OH transitions for the ``A'' and ``B'' components. The figures in which the lines are shown are indicated in the last column.}\label{sma_flux2}
\begin{center}
\begin{tabular}{llllllll}
  \hline
  \hline
  Molecule & Freq.  & Transition & $E_{\rm u}$ & $\mu^2S$ & \multicolumn{2}{c}{\underline{$F\Delta V^a$ (Jy~km~s$^{-1}$) [$\sigma$]}} & Figure\\
  & (GHz)   &            & (K)         & (D$^2$) &  \phantom{1}``A''& ``B''\\
  \hline
  CH$_2$CO & 220.178 & 11$_{1,11}$--10$_{1,10}$ & \phantom{1}76 & \phantom{1}65.4 & \phantom{$<$}3.98 [18$\sigma$]& 3.30 [35$\sigma$] & \ref{overview}\\
  \hline
  CH$_3$CHO-A$^b$ & 219.780 & 11$_{1,10}$--10$_{1,9}$ & 435 &  \phantom{1}66.4 & $<$0.22 & 0.43 [\phantom{1}7$\sigma$] & \ref{ap3}\\
  & 230.302 & 12$_{2,11}$--11$_{2,10}$ &  \phantom{1}81 &  \phantom{1}73.7 & $<$0.22 & 2.23 [24$\sigma$] & \ref{ap3}\\
  & 230.395 & 12$_{2,11}$--11$_{2,10}$ & 286 &  \phantom{1}74.1 & $<$0.22 & 1.01 [17$\sigma$] & \ref{ap3}\\
  & 230.438 & 12$_{0,12}$--11$_{0,11}$ & 440 &  \phantom{1}72.9 & $<$0.22 & 0.66 [\phantom{1}8$\sigma$]& \ref{ap3} \\
  & 231.330 & 12$_{5,8/7}$--11$_{5,7/6}$ & 129 & 125.5 & $<$0.22 & 2.66 [28$\sigma$] & \ref{ap3}\\
\hline
 CH$_3$CHO-E$^b$  & 231.357 & 12$_{3,9}$--11$_{3,8}$ & 299 & \phantom{1}71.4 & $<$0.22 & 1.32 [14$\sigma$] & \ref{ap4}\\
  & 231.363 & 12$_{5,7}$--11$_{5,6}$ & 129 & \phantom{1}62.7 & $<$0.22 & 2.44 [26$\sigma$] & \ref{ap4}\\
  & 231.369 & 12$_{5,8}$--11$_{5,7}$ & 129 & \phantom{1}62.7 & $<$0.22 & 1.94 [21$\sigma$] & \ref{ap4}\\
  \hline
  C$_2$H$_5$OH & 220.602 & 13$_{1,13}$--12$_{0,12}$ &  \phantom{1}74 &  \phantom{1}44.7 & \phantom{$<$}--$^c$ & 4.02 [34$\sigma$] & \ref{overview}, \ref{ch3c13n_spec}, \ref{ch2co_sma}\\
  \hline
\end{tabular}
\end{center}
$^aF\Delta V$ is determined over the full width of the emission coming from source ``A'' or ``B'', respectively. $^b$3$\sigma$ upper limits are calculated assuming that the CH$_3$CHO-A and E lines have the same line-widths as CH$_2$CO. $^c$Blended with the CH$_3^{13}$CN 12$_3$--11$_3$ and CH$_3$CN 12$_6$--11$_6$ transitions.
\end{table*}

\subsection{Maps and spectra}\label{maps}

\begin{figure*}\centering
  \resizebox{\hsize}{!}{\includegraphics{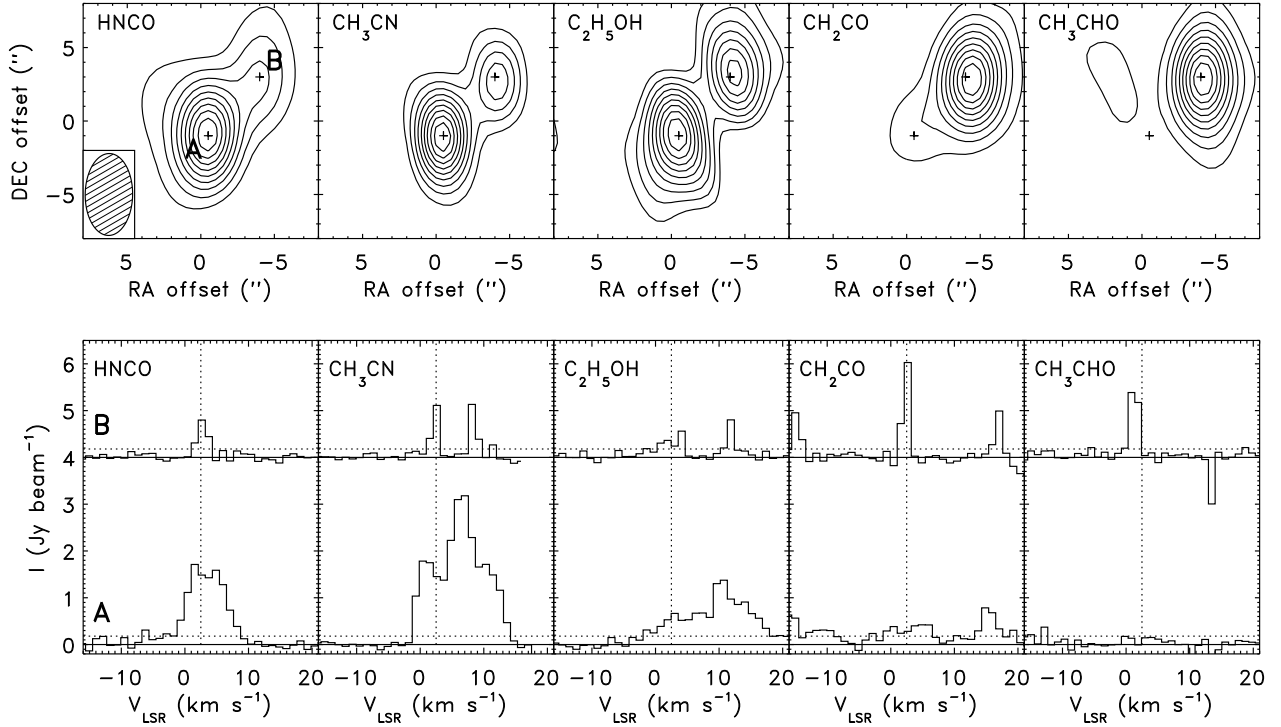}}
  \caption{Maps and spectra of the HNCO $10_{0,10}$--9$_{0,9}$,
    CH$_3$CN $12_0$--$11_0$, C$_2$H$_5$OH 13$_{1,13}$--12$_{0,12}$,
    CH$_2$CO 11$_{1,11}$--10$_{1,10}$ and CH$_3$CHO
    12$_{5,8/7,0}$--11$_{5,7/6,0}$ lines from -15--20~km~s$^{-1}$. The
    maps (upper panels) show the emission integrated over the width of
    the ``B'' component i.e. $\pm$1.0-1.5~km~s$^{-1}$ around the
    systemic velocity. The contour levels are at 10\%, 20\%, etc. of
    the peak in the maps. The peak integrated intensities are 10.3,
    3.5, 1.0, 3.2, 2.7~Jy~km~s$^{-1}$ for HNCO, CH$_3$CN,
    C$_2$H$_5$OH, CH$_2$CO and CH$_3$CHO, respectively. The size of
    the SMA beam is indicated in the lower left corner of the HNCO map
    for comparison. In the spectra (lower panels) the lower row
    indicates the spectra toward ``A'' and the upper row toward ``B''
    shifted in the vertical scale by 4~Jy~beam$^{-1}$. A dotted
    vertical line at $v_{\rm LSR}=$~2~km~s$^{-1}$ indicates the
    average systemic velocity for the compact emission toward ``A''
    and ``B''. The additional emission feature present at
    8~km~s$^{-1}$ next to CH$_3$CN $12_0$--11$_0$ is the CH$_3$CN
    $12_1$--11$_1$ transition and that next to CH$_2$CO
    11$_{1,11}$--10$_{1,10}$ is the HCOOCH$_3$
    17$_{2,15}$--$16_{4,12}$ transition. Note that small velocity
    differences are present between the sources and perhaps even
    between different species. Solid and dashed horizontal lines
    indicate the baseline and the 3$\sigma$ level per channel,
    respectively.}\label{overview}
\end{figure*}

\begin{figure}\centering
\includegraphics[width=9.2cm]{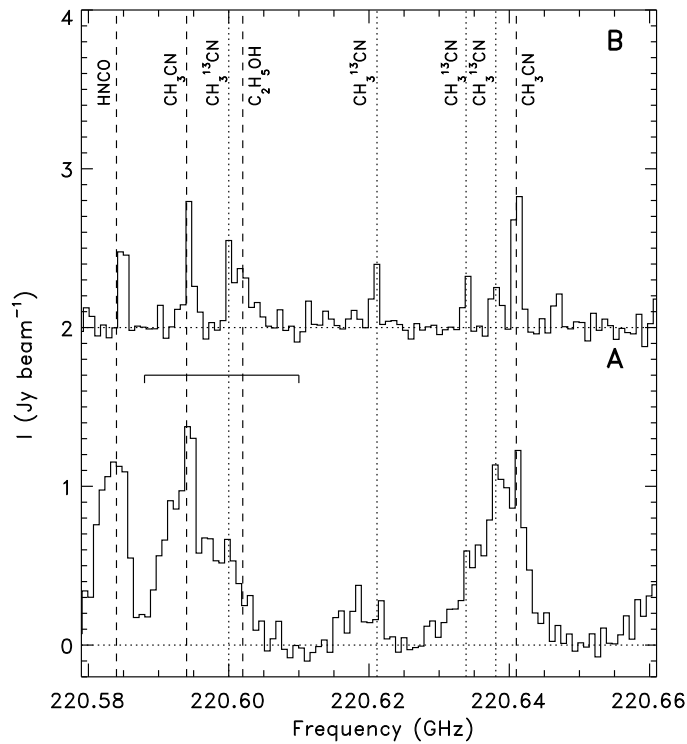}
\caption{SMA spectrum in the 220.580--220.66~GHz range for sources
  ``A'' (lower spectrum) and ``B'' (upper spectrum). The dotted lines
  indicate from left to right the positions of the 12$_3$--11$_3$,
  12$_2$--11$_2$, 12$_1$--11$_1$ and 12$_0$--11$_0$ CH$_3^{13}$CN
  transitions and the dashed lines the positions of lines for other
  species, in particular the 10$_{1,9}$--9$_{1,8}$ HNCO,
  12$_6$--11$_6$ and 12$_5$--11$_5$ CH$_3$CN and
  13$_{1,13}$--12$_{0,12}$ C$_2$H$_5$OH transitions. The frequency
  range shown in more detail in Fig.~\ref{ch2co_sma} is indicated with
  a bar above the spectrum of source ``A''. Note that the x-axis here
  is given in GHz and in Fig.~\ref{ch2co_sma} in km~s$^{-1}$, so that
  the spectra are reversed compared to each other. The horizontal
  dotted lines show the baseline.}\label{ch3c13n_spec}
\end{figure}

\begin{figure}\centering
  \includegraphics[width=9.2cm]{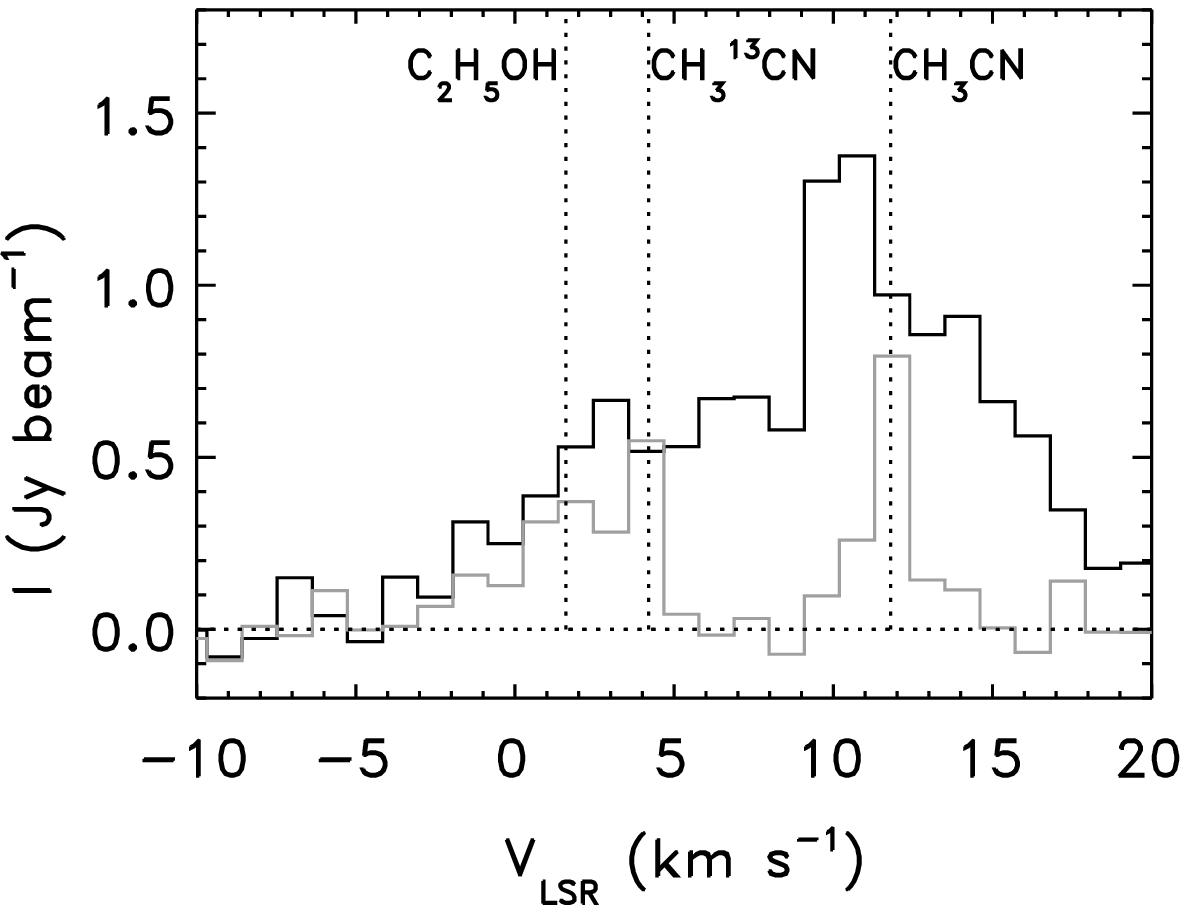}
  \caption{SMA spectrum of the C$_2$H$_5$OH 13$_{1,13}$--12$_{0,12}$
    line at 220.602~GHz (dotted line at 2.5~km~s$^{-1}$). The black
    line indicates the emission spectrum toward ``A'' and the grey
    line toward ``B''. The dotted vertical lines indicate the position
    of the CH$_3^{13}$CN 12$_3$--11$_3$ transition at 220.600~GHz
    (4~km~s$^{-1}$) and the CH$_3$CN 12$_6$--11$_6$ transition at
    220.594~GHz (12.5~km~s$^{-1}$) assuming their emission arises at
    $v_{\rm LSR}=$~2~km~s$^{-1}$. The horizontal dotted line shows the
    baseline.}\label{ch2co_sma}
\end{figure}

Figure~\ref{overview} displays the maps and selected spectra toward
the ``A'' and ``B'' components of IRAS~16923-2422, whereas the spectra
of additional lines are shown in Figures \ref{overview},
\ref{ch3c13n_spec}, \ref{ch2co_sma} and the online appendix~\ref{ap}
in Figures \ref{ap1}-\ref{ap4} (see also Tables~\ref{sma_flux} \&
\ref{sma_flux2}). The line identifications are based on comparison
with the CDMS\footnote{\tt http://www.astro.uni-koeln.de/cdms/},
JPL\footnote{\tt
  http://spec.jpl.nasa.gov/ftp/pub/catalog/catform.html} and
NIST\footnote{\tt
  http://physics.nist.gov/cgi-bin/micro/table5/start.pl} catalogs. The
assignments are assumed to be secure when no other species emits close
to the observed frequency. Note that for CH$_3$CHO the frequencies
from the NIST catalog have been used, because the data in the JPL
catalog are based on extrapolations of the line positions from lower
frequencies. These are shifted by a few MHz or more from the actual
measured line positions in the laboratory by \citet{kleiner1996}.

\begin{figure}\centering
  \includegraphics[width=9.2cm]{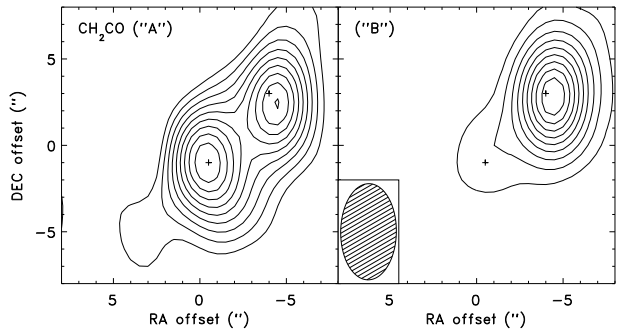}
  \caption{Maps for the CH$_2$CO 11$_{1,11}$--10$_{1,10}$ line
    integrated over the width of the ``A'' component of
    $\pm$4~km~s$^{-1}$ (left) and ``B'' component of
    $\pm$1-1.5~km~s$^{-1}$ (right) around the systemic velocity. The
    size of the SMA beam is shown in the map of the ``B'' component
    for comparison.}\label{ch2co_mom}
\end{figure}

From Figure~\ref{overview} it is clear that significant physical
differences exist between the regions where the emission from these
two sources arises. As previously noted by \citet{bottinelli2004} and
\citet{kuan2004} the spectra toward ``A'' show broad lines with a FWHM
of 8~km~s$^{-1}$, whereas the lines toward ``B'' are much narrower,
typically less than 2~km~s$^{-1}$ wide ($\sim$2~channels). The high
excitation lines of the complex organics peak at $V_{\rm
  LSR}=$1.5--2.5~km~s$^{-1}$ toward both sources. This is in contrast
to the systemic velocities of the larger scale envelope, e.g., traced
by HCN, at 3--4.5~km~s$^{-1}$ \citep[][]{takakuwa2007}. The maps
displayed in Fig.~\ref{overview} are obtained by integrating over the
width of the ``B'' component. Since the line-width in source ``A'' is
much larger and the peak intensity is much lower than that of source
``B'', only a fraction of the total flux of the CH$_2$CO line in
source ``A'' is shown in the maps in Fig.~\ref{overview} (see also
Table~\ref{sma_flux2}). When the CH$_2$CO line is integrated over the
line-width from source ``A'' of 8~km~s$^{-1}$ equally strong peaks for
both sources are detected as is illustrated in Fig.~\ref{ch2co_mom},
where the maps for CH$_2$CO are shown integrated over both velocity
ranges.

HNCO and CH$_3$CN show much weaker emission at the ``B'' position,
i.e., 8\% and 18\% respectively of that at the ``A'' position whereas
CH$_3$CHO is only seen toward the ``B''
position. Tables~\ref{sma_flux} and \ref{sma_flux2} list the
integrated line intensities for positions ``A'' and ``B''. When there
is no detection, 3$\sigma$ upper limits to the intensity are given
provided that there is no overlap with other species. Lines of
isotopologues were also searched to check the optical depth. No
HN$^{13}$CO lines were detected leading to a lower limit for
HNCO/HN$^{13}$CO of about 10. Since even high-mass objects show no
evidence for high optical depth \citep{zinchenko2000,bisschop2007a},
we assume for the remainder of this paper that the emission from HNCO
is optically thin. CH$_3^{13}$CN is clearly detected toward source
``B'' as is shown in Fig.~\ref{ch3c13n_spec} for the CH$_3^{13}$CN
branch. The CH$_3^{13}$CN transitions (see Fig.~\ref{ch3c13n_spec})
are strongly blended in source ``A''. However, it is probable that it
is present since there is emission toward source ``A'' that coincides
with the locations of the CH$_3^{13}$CN transitions. Due to the
line-blends, it is unfortunately not possible to determine accurate
integrated line-intensities for source ``A''. For source ``B'',
however, they can be determined. The CH$_3$CN/CH$_3^{13}$CN ratio is
$\sim$10 with a relatively large uncertainty on the fluxes of 40\% due
to the low signal-to-noise. The CH$_3$CN/CH$_3^{13}$CN ratio is much
lower than the $^{12}$C/$^{13}$C ratio of 77$\pm$7 determined by
\citet{wilson1994} for the local ISM and the estimated optical depth
is $\approx$2. The CH$_3$CN emission for ``A'' is probably optically
thick as well. A high optical depth toward IRAS~16293-2422 for
CH$_3$CN was also found by \citet{bottinelli2004}. Note that for high
optical depth the rotation temperature and number of molecules derived
from the rotation diagram will be upper and lower limits,
respectively.

The C$_2$H$_5$OH 13$_{1,13}$--12$_{0,12}$ transition is detected
toward source ``B'' shown in Figs~\ref{ch3c13n_spec} and
\ref{ch2co_sma}. It is strongly blended with the CH$_3^{13}$CN
12$_3$--11$_3$ transition at 220.600~GHz, and for source ``A'' also
with the CH$_3$CN 12$_{6}$--11$_6$ transition at 220.594~GHz. This
makes the estimate of the number of molecules from this line very
uncertain, even though it is detected and the identification of the
line seems secure for source ``B''.

\section{Analysis}
\label{an}
\subsection[Compact and extended emission]{Disentangling compact and
  extended emission: fits in the $(u,v)$-plane}\label{uvan}
One of the issues in dealing with interferometric observations of
protostars is that a significant fraction of the emission may be
resolved out, especially for low excitation transitions of molecules
also present in the large scale cold envelopes. Furthermore, fits in
the image plane suffer from the effects of the Fourier transformation
and image deconvolution, in particular for sparsely sampled data.

\begin{figure}\centering
  \includegraphics[width=9.2cm]{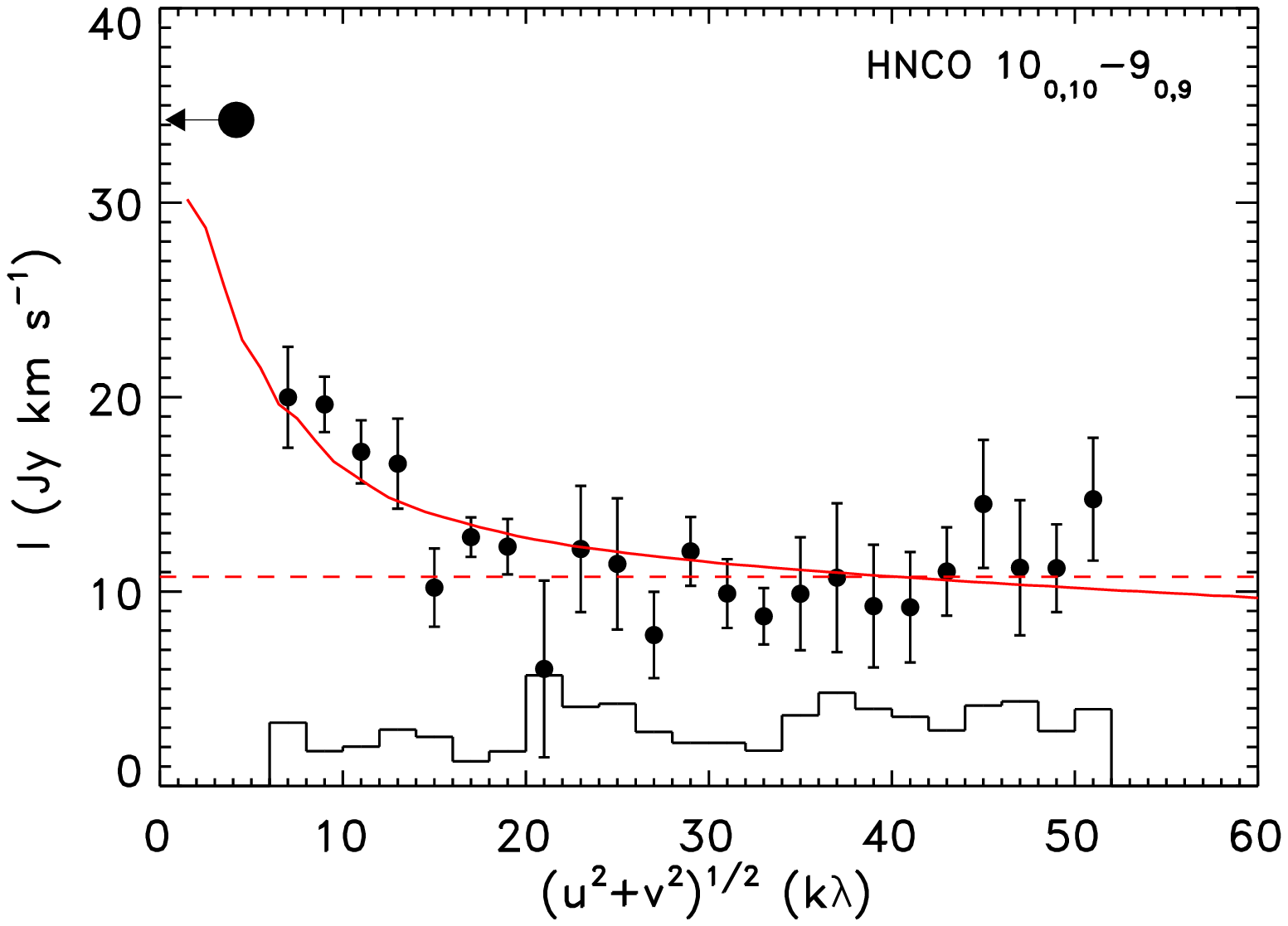}
  \caption{Visibility amplitudes of the emission for the HNCO
    10$_{0,10}$--9$_{0,9}$ transition at 219.798~GHz observed with the
    SMA as a function of the projected interferometric baseline length
    with the phase center taken at 1$''$ north of the ``A''
    component. The large filled circle near 35~km~s$^{-1}$ indicates
    the flux detected for the same line with the JCMT. The solid line
    represents a fit to the flux at different baselines using the
    \citet{schoier2002} model and assuming a ``jump'' in the abundance
    for HNCO at 90~K. The dashed line corresponds to the average
    visibility amplitude for baselines longer than 17~k$\lambda$. The
    histogram gives the expected amplitude for zero observed signal,
    i.e., in the absence of source emission.}\label{hnco_sma1}
\end{figure}

We have analyzed the data directly in the $(u,v)$-plane of the
interferometric observations to investigate the molecular emission as
function of baseline length, where longer baselines correspond to more
compact emission, and compared this to the single dish flux measured
by the JCMT. Figure~\ref{hnco_sma1} illustrates this for the lowest
excitation HNCO $10_{0,10}$--$9_{0,9}$ line at 219.798~GHz ($E_{\rm
  u}$~=~58~K). These observations show a break around 15~k$\lambda$
(2000~AU) with an almost constant amplitude at longer baselines,
implying that there is an additional compact (unresolved) component
contributing to the emission. This break is consistent with the
analysis of single dish data by \citet{schoier2002}, which gives an
abundance enhancement in the innermost envelope. For shorter baselines
the flux increases and the single dish flux with the JCMT is recovered
within the uncertainty. When the analysis is restricted to the longer
baselines, the emission can be compared for compact emission
exclusively without worrying about the presence of extended envelope
material. For all other lines and species the point source fits are
therefore made to baselines longer than 17~k$\lambda$ integrated over
the velocity range of each component. This results in the fluxes
listed in Tables~\ref{sma_flux} and \ref{sma_flux2}. For the HNCO
$10_{0,10}$--$9_{0,9}$ transition, about a third of the flux is
observed at the longest baselines, due to compact emission, whereas
the remaining two thirds can be attributed to the extended cold
envelope. This is the only species, however, where a significant
increase is seen for shorter baselines. Thus, most of the emission
from the remaining species is unresolved although colder extended
material is not excluded (see \S~\ref{results}).

\subsection{Rotational diagrams}
\label{rot}
When multiple transitions of a given molecule are observed spanning a
large range in excitation energies it is possible to infer the
physical properties of the emitting gas under various assumptions. The
rotational diagram method \citep[e.g.,][]{goldsmith1999} is often used
for molecules of different excitation energies. Under the assumption
that the emitting gas is optically thin a straightforward relationship
exists between the integrated intensities, the column density of the
molecule and its so-called ``rotational temperature''. At high densities
this will approach the kinetic temperature of the gas.

For single-dish observations one typically assumes that the source is
compact with respect to the beam and that a specific beam filling
factor (or source size) applies as a correction to the observed
intensities. For the interferometric observations presented here we
know that the source is unresolved at a specific angular
resolution. It is therefore more natural to work directly in units of
flux densities, so that the analysis is independent of an unknown
source size. For an optically thin source, ignoring background
radiation, the source brightness at a specific frequency, $I_{\rm
  \nu}$ is then given by:

\begin{equation}
  I_\nu = \frac{A_{\rm ul}N_{\rm u}^{\rm thin}hc}{4\pi\Delta V},
\end{equation}

\noindent where $A_{\rm ul}$ is the Einstein coefficient, $N_{\rm
  u}^{\rm thin}$ the column density of molecules in the upper energy
level and $\Delta V$ the emission line-width. Assuming the source
emits uniformly over a solid angle, $\Omega$, we can also express the
column density $N_{\rm u}^{\rm thin}$ as the total number of
molecules, $Y_{\rm u}^{\rm thin}=N_{\rm u}^{\rm thin} \Omega$. Since
the definition of brightness of a uniform source is $I_\nu =
{S_\nu}/{\Omega}$ where $S_\nu$ is the flux density, we can write:

\begin{equation}
  S_\nu = \frac{A_{\rm ul}Y_{\rm u}^{\rm thin}hc}{4\pi\Delta V}.
\end{equation}

\noindent Note that in contrast to the typical calculation for rotational
diagrams, the source size does not enter the equation here - only the
total number of molecules. The observed quantity from this expression
is the integrated line intensity, $S_\nu \Delta V$, typically measured
in Jy~km~s$^{-1}$.

From this expression it is straightforward to isolate the number of
molecules in the energy level, $Y_{\rm u}^{\rm thin}$, and through the
Boltzmann equation relate it to the total number of molecules, $Y_{\rm
  T}$:
\begin{equation}
  \frac{Y_{\rm u}^{\rm thin}}{g_{\rm u}} = \frac{Y_{\rm T}}{Q(T_{\rm rot})} e^{-E_{\rm u}/T_{\rm rot}}\ ,
\end{equation} 

\noindent where $g_{\rm u}$ is the degeneracy of the upper energy
level, $Q$ the molecular partition function, $T_{\rm rot}$ the
rotational temperature and $E_{\rm u}$ the energy of the upper energy
level. As in the usual rotational diagram analysis, one can then plot
${Y_{\rm u}^{\rm thin}}/{g_{\rm u}}$ versus $E_{\rm u}$, and derive
the rotational temperature, $T_{\rm rot}$, from the slope and the
total number of molecules, $Y_{\rm T}$, from the interpolation to
$E_{\rm u}$~=~0~K. Column densities can be derived for interferometric
observations but as for the single-dish observations an assumption for
source size has to be made. 

In this way rotational diagrams were fit for all molecules with a
sufficient number of detected lines. The results are summarized in
Table~\ref{rot_fit}. For some species no rotational temperature could
be determined, since too few lines were detected. In that case $Y_{\rm
  T}$ is derived assuming that these species have the same rotational
temperature as other chemically related species in the same source. In
\S~\ref{results} we discuss these results in detail for each molecule.

\section{Results}\label{results}
\subsection{HNCO and CH$_3$CN}
\label{hnco_res}

\begin{table*}\begin{center}
    \caption{Rotational temperature, $T_{\rm rot}$, total number of
      molecules, $Y_{\rm T}$, and abundance with respect to HNCO,
      $x_{\rm HNCO}$. }\label{rot_fit}
\begin{tabular}{l|lll|lll}
  \hline
  \hline
  Molecule & \multicolumn{3}{c|}{Source ``A''} & \multicolumn{3}{c}{Source ``B''}\\
  & \phantom{$>$}$T_{\rm rot}$ & \phantom{$>$}$Y_{\rm T}^a$ & \phantom{$>$}$x_{\rm HNCO}$ & \phantom{$>$}$T_{\rm rot}$ & \phantom{$>$}$Y_{\rm T}^a$ & $x_{\rm HNCO}$\\
  & \phantom{$>$}(K)        & \phantom{$>$}(mol.)  &      & \phantom{$>$}(K)        & \phantom{$>$}(mol.) &\\
  \hline
  HNCO     & \phantom{$>$}277$\pm$76 & \phantom{$>$}8.9(14)   & \phantom{$>$}1.0         & \phantom{$>$}237$\pm$56 & \phantom{$>$}7.5(13) & \phantom{1}1.0\\
  CH$_3$CN$^b$ & $<$372$\pm$110 & $\geq$5.6(14)    & $>$0.6        & $<$390$\pm$99 & \phantom{$>$}1.1(15)$^c$ & 14\\
  CH$_3^{13}$CN & \phantom{$>$}---   & \phantom{$>$}---  & \phantom{$>$}--- & \phantom{$>$}261$\pm$485 & \phantom{$>$}1.4(13)$^c$ & \phantom{1}0.19\\
  CH$_2$CO$^d$ &\phantom{$>$}---        & \phantom{$>$}5.6(14) & \phantom{$>$}0.6       & \phantom{$>$}---        & \phantom{$>$}4.7(14) & \phantom{1}6.3\\
  CH$_3$CHO & \phantom{$>$}---       & $<$2.2(13)$^e$      &   $<$0.02$^e$       & \phantom{$>$}250$\pm$49 & \phantom{$>$}3.2(14) & \phantom{1}4.3\\
  C$_2$H$_5$OH$^d$ & \phantom{$>$}---    & \phantom{$>$}---  & \phantom{$>$}---              & \phantom{$>$}---        & \phantom{$>$}7.1(14) & \phantom{1}9.5\\
  \hline
\end{tabular}
\end{center}
$^a$The notation $y(z)$
stands for $y$ $\times$10$^z$.$^bT_{\rm rot}$ for CH$_3$CN in both sources is calculated assuming the main
isotope is optically thin, and is therefore an upper limit (see \S~\ref{rot}), whereas $Y_{\rm T}$ is
a lower limit. $^cY_{\rm T}$ for CH$_3$CN and CH$_3^{13}$CN in source ``B'' has been determined from the CH$_3^{13}$CN lines assuming a $^{12}$C/$^{13}$C ratio of 77 \citep{wilson1994}. $^dY_{\rm T}$ for CH$_2$CO and
C$_2$H$_5$OH in source ``B'' are derived assuming they have the same rotational temperature as
CH$_3$CHO (see
Table~\ref{sma_flux2}). $^e$The upper limit is derived for CH$_3$CHO toward source ``A'' with the assumption that the rotational temperature is identical to that in source ``B''.

\end{table*}

\begin{figure*}\centering
  \resizebox{\hsize}{!}{\includegraphics[width=6cm]{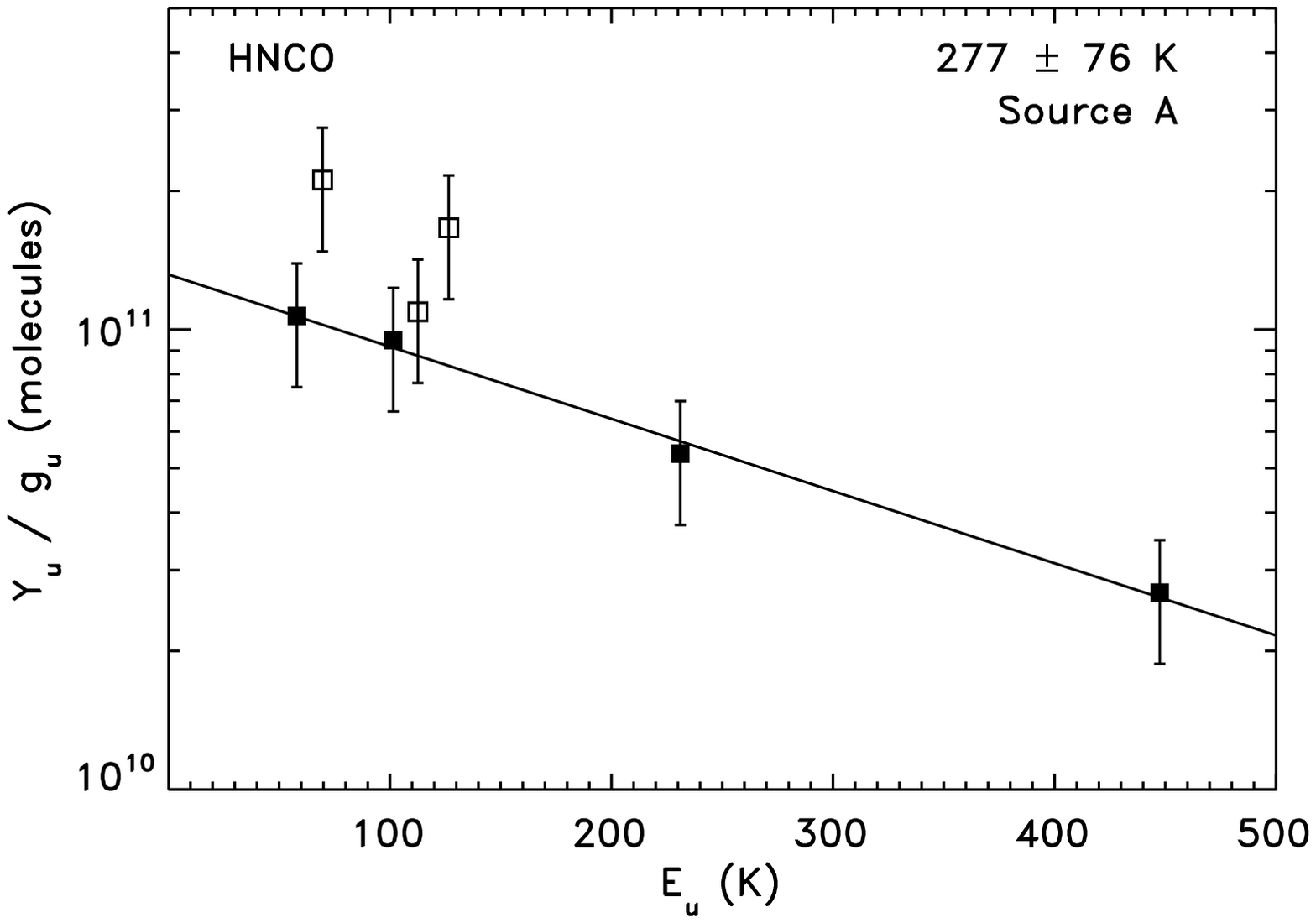}\includegraphics[width=6cm]{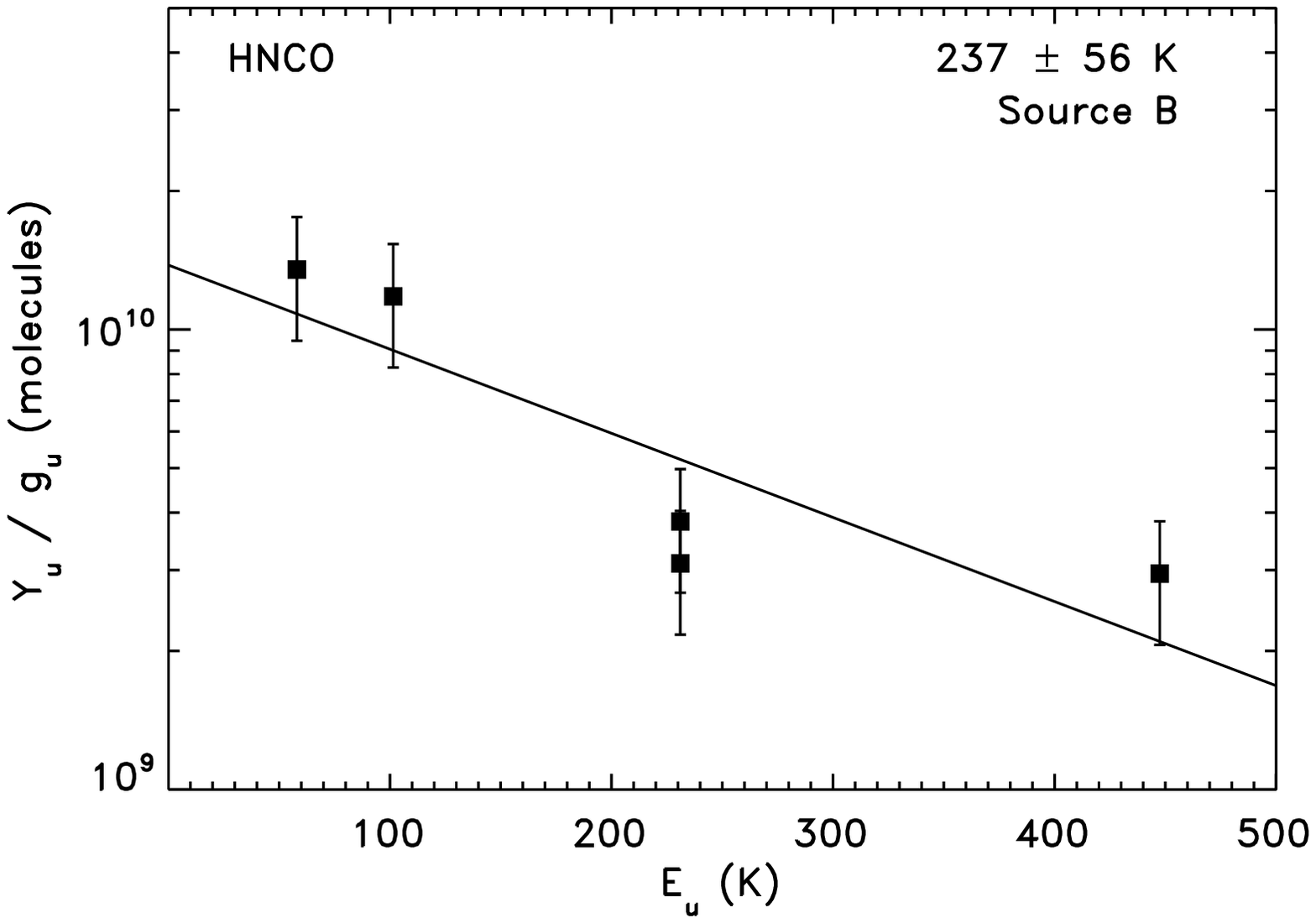}}
  \caption{Rotational diagram for HNCO transitions detected with the SMA
    shown as filled squares ($\blacksquare$) toward sources ``A''
    (left) and ``B'' (right). Emission lines detected with the single
    dish telescopes JCMT and CSO by \citet{dishoeck1995} are indicated
    with open squares ($\Box$), but are not used for the
    fit. Uncertainties of 30\% are over-plotted on the
    data.}\label{hnco_sma_rot}
\end{figure*}

The rotational diagrams for HNCO are shown in Fig.~\ref{hnco_sma_rot}
and the resulting $T_{\rm rot}$ and $Y_{\rm T}$ are given in
Table~\ref{rot_fit} for source ``A'' and ``B''. For source ``A'' the
available single dish fluxes from the literature by
\citet{dishoeck1995} have been over-plotted, but have not been used
for the fit. The single dish covers both components, but because the
flux from source ``A'' is significantly stronger than ``B'' (see
\S~\ref{maps}), we assume that the single dish flux arises from source
``A''. The single dish fluxes are slightly larger than the fluxes from
the SMA, also when the additional flux from the ``B'' source is taken
into account. The rotational temperatures are the same for source
``A'' and ``B'', 277$\pm$76 and 237$\pm$56~K respectively, and much
higher than what is detected by \citet{dishoeck1995} of
135$\pm$40~K. This discrepancy is due to the fact that
\citet{dishoeck1995} included lines in their fit which have rotational
temperatures of only up to $\sim$130~K, which results in a lower
rotational temperature. Additionally, it is possible that extra flux
in the single dish observations is due to extended emission from the
larger scale envelope (see \S~\ref{uvan}). The lines detected here
range from 58~K to 447~K and have stronger constraints on the
rotational temperature of compact HNCO emission. It is important to
note that IRAS~16293-2422 emits at mid-infrared wavelengths
\citep{jorgensen2005}, and thus infrared pumping of HNCO is
possible. In that case the rotational temperature of HNCO is higher
than the gas kinetic temperature - and thus than rotational
temperatures determined for other molecules. When the model by
\citet{schoier2002} is used to fit the $(u,v)$-plot (see
Fig.~\ref{hnco_sma1}), the abundance of HNCO with respect to H$_2$ can
be determined for the outer and inner envelope to be
6.0$\times$10$^{-11}$ and 4.0$\times$10$^{-9}$ assuming there is a
jump in the abundance at 90~K. The inner abundance is somewhat
uncertain, because the exact source size for the compact emission is
not known.

\begin{figure*}\centering
  \resizebox{\hsize}{!}{\includegraphics[width=6cm]{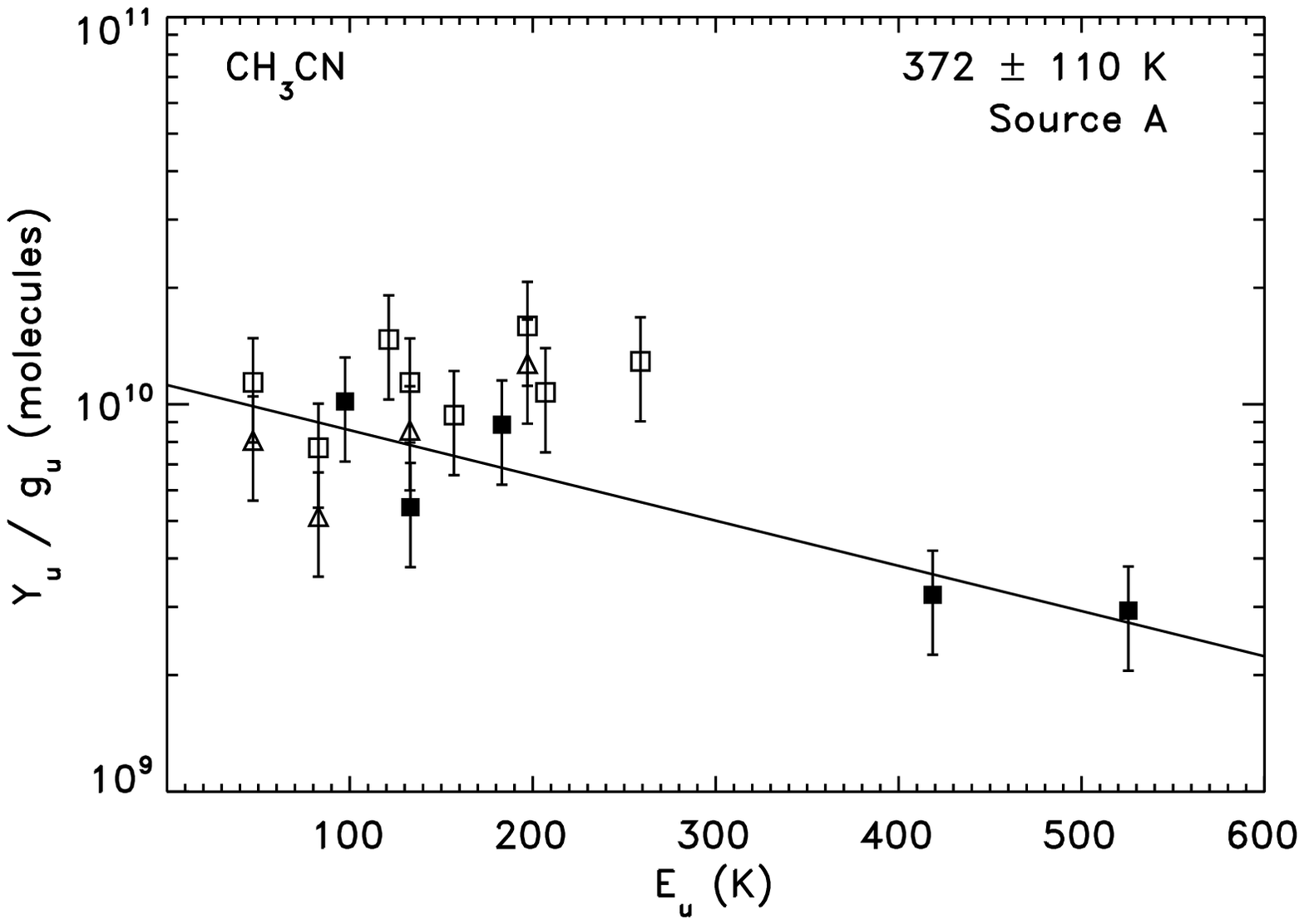}\includegraphics[width=6cm]{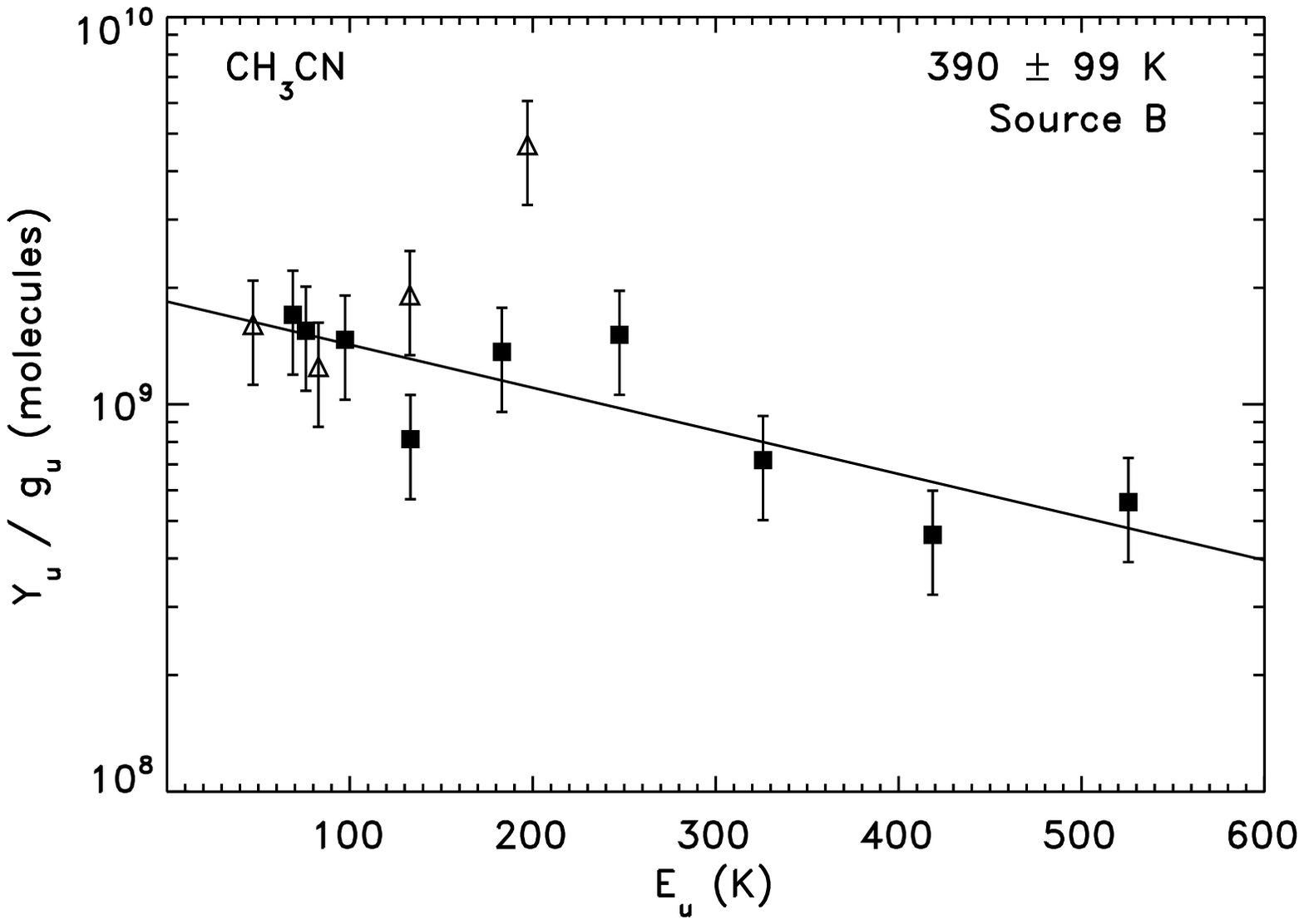}}
  \caption{Rotational diagram for CH$_3$CN transitions detected with the
    SMA shown as filled squares ($\blacksquare$) toward
    IRAS~16293-2422 for source ``A'' (left panel) and source ``B''
    (right panel). Emission lines detected with the JCMT observations
    by \citet{dishoeck1995} and the IRAM 30~m observations by
    \citet{cazaux2003} are shown for source ``A'' with open squares
    ($\Box$). Emission lines detected by \citet{bottinelli2004} with
    the Plateau de Bure interferometer are shown for both sources with
    open triangles ($\bigtriangleup$). Only the SMA data studied here
    is used for the fits. Uncertainties of 30\% are over-plotted on
    the data.}\label{ch3cn_rotA}
\end{figure*}

\begin{figure}[htp]\centering
  \includegraphics[width=9.2cm]{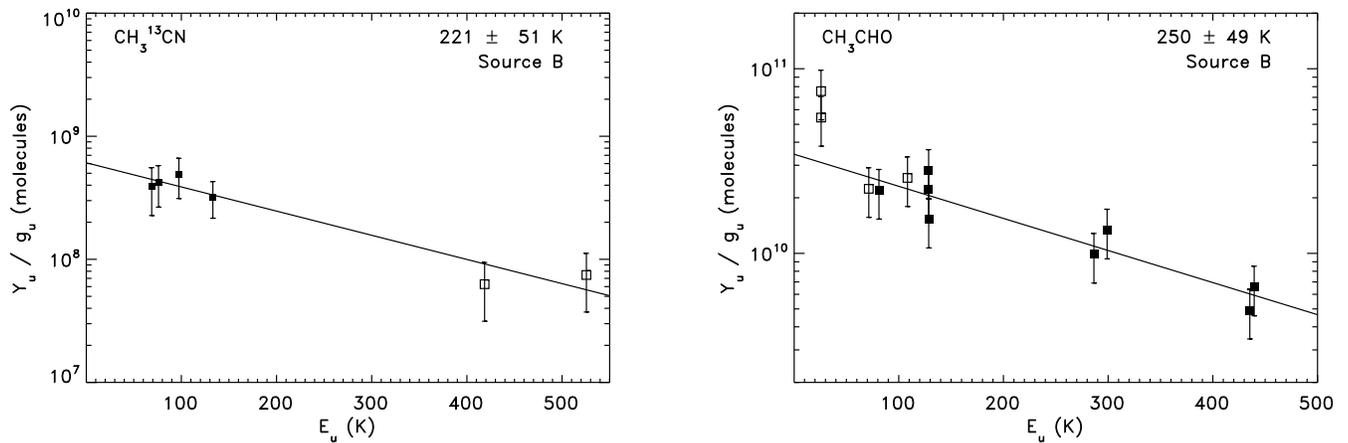}
  \caption{Rotational diagram for CH$_3^{13}$CN emission detected with
    the SMA toward source IRAS~16293-2422B shown as filled squares
    ($\blacksquare$) and high excitation lines of the main isotope
    CH$_3$CN corrected for the $^{12}$C/$^{13}$C isotopic ratio are
    shown as open squares ($\Box$). Uncertainties are over-plotted on
    the data.}\label{ch3-13cn}
\end{figure}

For CH$_3$CN rotational diagrams are constructed for both sources (see
Fig.~\ref{ch3cn_rotA}). Less lines are usable for source ``A'' due to
the larger line-width and line-blending. The single dish detections by
\citet{dishoeck1995} and \citet{cazaux2003} are over-plotted on the
data for source ``A'' and are consistent with the SMA data within the
uncertainties. For both sources the interferometric observations by
\citet{bottinelli2004} with the Plateau de Bure interferometer have
been over-plotted. These are also consistent with the SMA data, except
for the 6$_5$--5$_5$ transition at 110.329~GHz with $E_{\rm u}=$~197~K
which is clearly too high for source ``B''. This may be due to line
blending with the CH$_3^{13}$CN 6$_0$--5$_0$ transition, which has a
larger line strength than the main isotope. It is thus plausible that
it contributes significantly to the line-flux. The resulting
rotational temperatures are the same within the uncertainties for
source ``A'' and ``B'' (see Table~\ref{rot_fit}). The transitions of
the isotopomer CH$_3^{13}$CN are unfortunately strongly blended in
source ``A'' so that the determination of the optical depth is not
possible. From the detections of lines for CH$_3^{13}$CN for source
``B'' it is clear that the emission is optically thick and therefore
that the resulting $T_{\rm rot}$ and $Y_{\rm T}$ from the rotational
diagram of the main isotope are over- and underestimated, respectively
(see Sect.~\ref{maps}). A rotational diagram has been constructed for
source ``B'' from the four CH$_3^{13}$CN lines and is shown in
Fig.~\ref{ch3-13cn}. The results of this fit are not well constrained
because they are relatively close in excitation energy. However, the
line intensities of the two highest excitation CH$_3$CN transitions,
which are likely optically thin, can be added. When they are corrected
for the $^{12}$C/$^{13}$C isotopic ratio of 77 \citet{wilson1994},
they are consistent with the fit and thereby decrease the uncertainty
from 261$\pm$485~K to 221$\pm$51~K (see Fig.~\ref{ch3-13cn}). Using
the value of $T_{\rm rot}=$261~K we can derive a number of molecules
for CH$_3$CN for source B (see Table~\ref{rot_fit}). For CH$_3$CN no
extended emission can be inferred from the $(u,v)$-data. The lowest
detected emission line has $E_{\rm u}$~=~69~K in our SMA data. Hence
we cannot exclude the presence of extended emission for lines with
even lower excitation energies.

\subsection{CH$_2$CO, CH$_3$CHO and C$_2$H$_5$OH}\label{ch2co}
CH$_2$CO has previously been detected by \citet{dishoeck1995} who
found only very weak lines with the JCMT and CSO. Their derived
rotational temperature is 35~K and column density
4$\times$10$^{13}$~cm$^{-2}$. Furthermore, \citet{kuan2004} detected
the CH$_2$CO 17$_{2,15}$--16$_{2,14}$ transition at 343.694~GHz
($E_{\rm u}$~=~200~K) only toward source ``B'', suggesting that hot
CH$_2$CO is also present. The 11$_{1,11}$--10$_{1,10}$ transition
($E_{\rm u}$~=~76~K) detected here has a similar strength in both
sources. A comparison with the single dish fluxes from
\citet{dishoeck1995} shows that the SMA flux is lower by about a
factor of 2. This may indicate that large scale emission from CH$_2$CO
is resolved out with the SMA, consistent with the low rotational
temperatures found in single-dish data
\citep{dishoeck1995}. Unfortunately, it is difficult to derive the
presence of extended emission from the $(u,v)$-visibility plots for
the SMA observations, because the line strengths in source ``A'' and
``B'' are so similar. Future higher angular resolution observations
are needed to resolve this issue. Since only one line is detected for
both CH$_2$CO and C$_2$H$_5$OH, it is not possible to determine
rotational temperatures. $Y_{\rm T}$ can however be estimated when one
assumes that CH$_2$CO and C$_2$H$_5$OH have the same rotational
temperature as CH$_3$CHO.

\begin{figure}\centering
  \includegraphics[width=9.2cm]{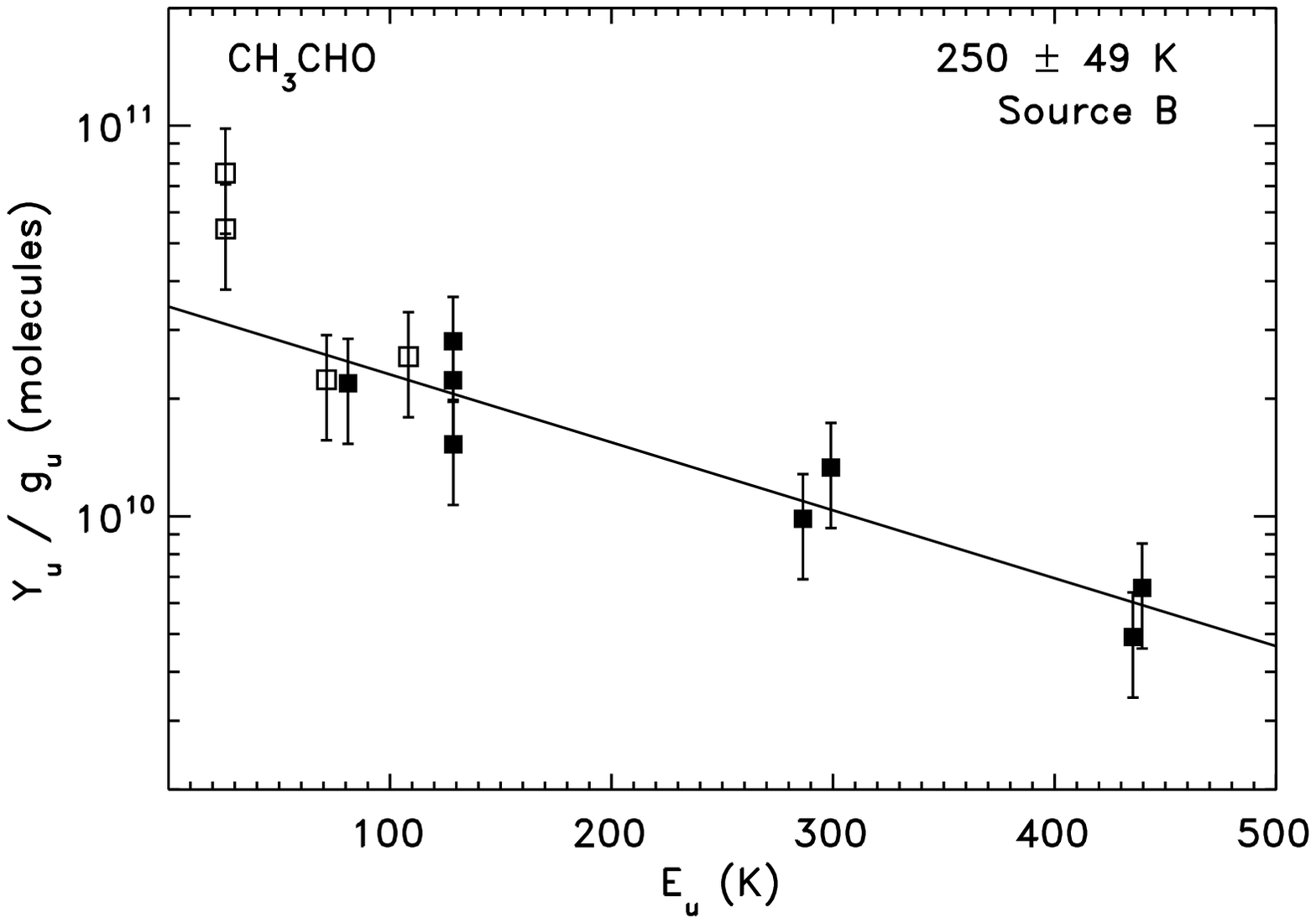}
  \caption{Rotational diagram for CH$_3$CHO emission detected with the
    SMA toward source IRAS~16293-2422B shown as filled squares
    ($\blacksquare$). The single dish data by \citet{cazaux2003} are
    shown as open squares ($\Box$), but are not used for the
    fit. Uncertainties of 30\% are over-plotted on the
    data.}\label{ch3cho_sma}
\end{figure}

Figure~\ref{ch3cho_sma} shows the rotational diagram of the CH$_3$CHO
lines detected with the SMA and the resulting values for $T_{\rm
  rot}$, $Y_{\rm T}$ and the abundance $x_{\rm HNCO}$ with respect to
HNCO are given in Table~\ref{rot_fit}. Both CH$_3$CHO-A and
CH$_3$CHO-E lines have been included in the fit with the assumption
that their abundance ratio is unity and they both trace the same
temperature gas. However, the result is not strongly influenced by
this assumption since the fit to both CH$_3$CHO-A and E lines is
within the uncertainties identical to the fit obtained to CH$_3$CHO-A
lines only. The single dish data by \citet{cazaux2003} are
over-plotted on the rotational diagram but are not used for the
fit. They are consistent with the SMA fluxes for $E_{\rm u}>$50~K. The
rotational temperature calculated from the SMA data is
281$\pm$47~K. \citet{cazaux2003} find a rotational temperature of
$\leq$40~K, and furthermore their lower excitation lines lie
significantly above the fit to the rotational diagram. This implies
that the emission with $E_{\rm u}<$30~K originates from cold extended
material. This is also consistent with interferometric observations of
CH$_3$CHO in the high-mass star forming region Sgr B2(N)
\citep{liu2005}, where the CH$_3$CHO emission is largely extended. The
limit to the CH$_3$CHO/CH$_3$OH abundance ratio in source ``A'' is an
order of magnitude lower than the detection in source ``B'', resulting
both from a lower CH$_3$CHO and a higher CH$_3$OH column.

Previously, \citet{dishoeck1995} suggested that C$_2$H$_5$OH could be
present in the spectra of this source, but were not able to make a
firm identification based on strong line-blends. \citet{huang2005}
made the first conclusive detection of compact emission for
C$_2$H$_5$OH toward source ``A'' for the 20$_{3,17}$--19$_{3,16}$
transition at 354.757~GHz with $E_{\rm u}$~=~249~K. The detection here
seems secure, but is blended with other lines (see \S~\ref{maps}). The
previous detection by \citet{huang2005} gives additional weight to the
assignment of C$_2$H$_5$OH.

The emission of CH$_2$CO, CH$_3$CHO and C$_2$H$_5$OH behaves spatially
very differently, also with respect to HNCO and CH$_3$CN. CH$_2$CO has
emission that is similar in strength for source ``A'' and ``B'' and
for C$_2$H$_5$OH this also appears to be the case even though due to
line blending it is not possible to determine this with
certainty. However, the limit on $Y_{\rm T}$ for CH$_3$CHO is an order
of magnitude lower than the detection toward source ``B''. The
emission from CH$_2$CO and C$_2$H$_5$OH may arise from similar
regions, but more transitions are required for a determination of the
rotational temperatures.
 
\section{Discussion}\label{disc}

In this section the chemical implications of the observations are
discussed. To compare the abundances of the species studied here with
other sources, they are determined relative to the CH$_3$OH
observations by \citet{kuan2004} with the SMA (see
Table~\ref{comp}). The value of $Y_{\rm T}$ has been calculated for
CH$_3$OH from the integrated flux given by \citet{kuan2004} assuming
that the emission comes from a compact region with an average
temperature of 200~K, a typical value found for the species in this
study. The abundances are presented for individual components and the
sum of ``A + B'' from the SMA data as well as previously published
single dish results for IRAS~16293-2422 \citep{dishoeck1995}, the
high-mass Orion~hot~core \citep{sutton1985} and the average over a
survey of seven high-mass YSOs \citep{bisschop2007a}. CH$_3$OH is
taken as a reference molecule because it is one of the most abundant
grain-surface products. Ideally, one would like to calculate the
molecular abundances with respect to H$_2$ for the most direct
comparison to chemical models. However, no accurate measure of the
number of H$_2$ molecules exist: the dust continuum emission at
(sub)millimeter wavelengths observed by interferometers is a
combination of emission from both the circum-binary envelope and the
emission from each of the compact disks around source ``A'' and ``B''
\citep[e.g.,][]{schoier2004}. Likewise, models for the larger-scale
circum-binary envelope \citep[e.g.,][]{ceccarelli2000,schoier2002}
have typically been 1-dimensional and thus not meaningful for
comparison of the relative abundances toward source ``A'' and
``B''. The uncertainties in the number of molecules for CH$_3$OH lead
to a possible error of -40\% to +70\% in the absolute abundance ratios
in Table~\ref{comp}. For all molecules at the same position (``A'' or
``B'') it is only $\sim$30\%.

\subsection{HNCO and CH$_3$CN}

It is striking that even though the line-widths of HNCO and CH$_3$CN
are significantly different between the ``A'' and ``B'' components,
their excitation and distribution is similar. When optical depth is
taken into account it is very probable that both species are present
in a similar environment. 

Since HNCO and CH$_3$CN are nitrogen-bearing species their relative
abundances compared to CH$_3$OH are not expected to be
constant. However, it is interesting to note that the abundance ratios
found for the low-mass IRAS~16293-2422 source as a whole (``A'' +
``B'') are very similar to the Orion~hot~core \citep{sutton1985} and a
high-mass survey \citep{bisschop2007a}. Specifically, HNCO/CH$_3$OH is
0.02 for ``A'' and CH$_3$CN/CH$_3$OH is $>$0.02, to be compared with
0.01--0.02 and 0.04--0.08 for the high-mass sources, respectively. In
contrast, for HNCO the differences between the two IRAS~16293-2422
sources ``A'' and ``B'' are larger than those with the high-mass
sources. This suggests that both species are formed through the same
mechanisms independent of cloud mass and luminosity, and that other
factors cause the differences between individual objects.

The most likely formation mechanism for HNCO is OCN$^-$ evaporation
from grains \citep{allamandola1999,broekhuizen2004}, whereas CH$_3$CN
may partly form on the surfaces of grains from reactions of CH$_3$
with CN as well as in the gas phase from HCN with CH$_3^+$
\citep{garrod2008}. One possible explanation for the large abundance
difference between ``A'' and ``B'' is that the initial solid state
abundances of nitrogen-bearing species such as OCN$^{-}$ and NH$_3$
were higher in ``A'' than in ``B'', which resulted in higher gas phase
HNCO abundance after evaporation. This may explain why the spatial
distribution of HNCO is so different also from the oxygen-bearing
species. Indeed the ice abundances of nitrogen-bearing species on
grains vary strongly between different objects as is exemplified by
the case of OCN$^-$ for which variations of about an order of
magnitude have been observed for sources that are only 400~AU apart
\citep{broekhuizen2005a}, similar to the distance between sources
``A'' and ``B''. Additionally, significant variations in the NH$_3$
ice abundance are seen in low-mass YSOs \citep[Bottinelli et al., in
prep.;][]{bottinelli2007b}. Additionally, the gas phase abundance
  of HC$^{15}$N is also almost an order of magnitude higher for source
  ``A'' compared to ``B'' \citet{kuan2004}, which suggests that the
  abundance of CH$_3$CN may also be enhanced in the gas phase.

\subsection{CH$_2$CO, CH$_3$CHO and C$_2$H$_5$OH}

\begin{table*}\begin{center}
    \caption{Comparison of IRAS~16293-2422 the abundances for source
      ``A'' and ``B'' normalized to the number of molecules of
      CH$_3$OH with those of other high-mass sources. }\label{comp}
\begin{tabular}{llllllll}
  \hline
  \hline
  Molecule$^a$ & ``A'' & ``B'' & ``A'' + ``B'' & Single Dish$^b$ & Orion KL hot core$^c$& High-mass survey$^d$\\
  \hline
  HNCO & \phantom{$<$}0.03 & 0.005 & \phantom{$<$}0.02                & \phantom{$<$}0.04 & 0.01 & \phantom{$<$}0.02\\
  CH$_3$CN & $>$0.02 & 0.07 & $>$0.02             & \phantom{$<$}0.04 & 0.04 & \phantom{$<$}0.08\\
  CH$_2$CO & \phantom{$<$}0.02 & 0.03 & \phantom{$<$}0.02                & \phantom{$<$}0.04 & 0.01 & $<$0.005$^e$\\
  CH$_3$CHO & $<$0.0007 & 0.02  & $<$0.007             & $<$0.02 & -- & $<$0.003$^e$\\
  C$_2$H$_5$OH & \phantom{$<$}---  & 0.01 & \phantom{$<$}---       & $\leq$0.05 & --& \phantom{$<$}0.025\\
  H$_2$CO  & \phantom{$<$}---  & ---  & \phantom{$<$}---             & \phantom{$<$}0.24 & 0.1 & \phantom{$<$}0.22\\
  \hline
\end{tabular}
\end{center}
$^a$From \citet{kuan2004}, assuming the flux arises from a compact
region with an excitation temperature of 200~K, $^b$\citet{dishoeck1995}, $^c$\citet{sutton1985}, $^d$\citet{bisschop2007a}, $^e$limits determined from highest excitation lines detected for CH$_2$CO and CH$_3$CHO in the high-mass YSO survey by \citet{bisschop2007a} assuming the emission comes from hot gas with $T_{\rm rot}=$~200~K. Since much lower rotational temperatures have actually been determined the values indicate upper limits only.
\end{table*}

A comparison of the abundances of CH$_2$CO, CH$_3$CHO and C$_2$H$_5$OH
with respect to CH$_3$OH in hot gas with the SMA \citep{kuan2004} is
given in Table~\ref{comp}. The hot CH$_2$CO and C$_2$H$_5$OH
abundances relative to CH$_3$OH are similar to those found in
high-mass sources. In contrast, the CH$_3$CHO abundance differs about
a factor of $\sim$30 between sources ``A'' and ``B''. This difference
arises from the fact that the limit on the number of molecules for
CH$_3$CHO is an order of magnitude lower toward source ``A'' compared
to ``B'', whereas the number of molecules for CH$_3$OH is a factor of
2--3 higher toward source ``A''.

\citet{tielens1997} explained the presence of these molecules through
successive hydrogenation of HCCO, where complete hydrogenation leads
to the formation of the most hydrogen-rich species C$_2$H$_5$OH and
incomplete hydrogenation to one of the intermediate species CH$_2$CO
and CH$_3$CHO. If hydrogenation of HCCO is the sole formation
mechanism for all species there are three possible outcomes, (i)
hydrogenation is complete, (ii) hydrogenation is incomplete or (iii)
CH$_3$CHO is very efficiently converted to molecules other than
C$_2$H$_5$OH at high temperatures in the ice or right after
evaporation from the grain surface. For scenario (i) behavior such as
observed for high-mass sources is expected with only C$_2$H$_5$OH seen
in the hot gas, since CH$_2$CO and CH$_3$CHO are fully converted to
C$_2$H$_5$OH. However, the detection of hot compact emission for
CH$_2$CO and CH$_3$CHO implies that this cannot be the case. Thus it
is possible that hydrogenation is incomplete (scenario ii). As
demonstrated in the experiments by \citet{bisschop2007c}, the relative
abundances of C$_2$H$_5$OH and CH$_3$CHO depend on time or rather on
the total number of H-atoms the ice is exposed to. Additionally, the
ice thickness affects the formation: for ices of a few monolayers
thick some CH$_3$CHO molecules can be hidden from impinging H-atoms
and this effect will depend strongly on ice morphology, such as
porosity and the presence of other species in the ice. It then becomes
difficult to predict the exact ratio between the relative abundances
of CH$_2$CO, CH$_3$CHO and C$_2$H$_5$OH. This ratio may vary strongly
per source.

One way to test scenario (ii) is to compare the abundance ratios
C$_2$H$_5$OH/CH$_3$CHO and CH$_3$OH/H$_2$CO. These ratios are expected
to be sensitive to the same ice morphology when grain-surface
hydrogenation is the main formation mechanism. This is not an
unreasonable assumption since H$_2$CO and CH$_3$OH were found to form
through grain-surface hydrogenation from HCO in laboratory experiments
\citep{watanabe2004,fuchs2007}. The CH$_3$OH/H$_2$CO abundance ratio
is $\sim$5 and in favor of CH$_3$OH, in contrast to 0.5 for the
C$_2$H$_5$OH/CH$_3$CHO ratio toward source ``B'' which is in favor of
CH$_3$CHO. It thus seems clear that there must be other processes than
just the simple hydrogenation involved. However, since both the
CH$_2$CO/CH$_3$OH ratio and C$_2$H$_5$OH/CH$_3$OH ratios are rather
similar between different sources, it appears that it is mainly the
chemistry of CH$_3$CHO that is not well understood.

Alternatively, CH$_3$CHO may be rapidly destroyed in the solid state
or in the gas phase right after evaporation (scenario iii), while
CH$_2$CO and C$_2$H$_5$OH are more stable. If this is correct, the
determining factor for CH$_3$CHO destruction is absent or less strong
for source ``B'' compared to ``A''. This could be an effect of e.g.,
UV-photolysis, although there is no a priori reason to expect very
different photodissociation rates for these three species. The more
quiescent nature of source ``B'' compared to ``A'' suggested by their
different line widths could also play a role. Temperature effects are
likely not important, since CH$_3$CHO has been detected with very high
temperatures toward source ``B''. Alternatively, CH$_3$CHO and
C$_2$H$_5$OH are not directly related. Further searches for CH$_3$CHO
abundance variations compared to C$_2$H$_5$OH and CH$_3$OH are needed
to elucidate its chemistry.

\section{Summary and conclusions}\label{sum}

We have performed an interferometric study of the complex organic
species HNCO, CH$_3$CN, CH$_2$CO, CH$_3$CHO and C$_2$H$_5$OH toward
the low-mass protostar IRAS~16293-2422 with the SMA. Previously
published data from \citet{kuan2004} are used to determine abundances
relative to CH$_3$OH. The main conclusions of this work are:

\begin{itemize}
\item The emission from both HNCO and CH$_3$CN is compact and is seen
  toward both sources in the binary, with only 10--20\% arising from
  source ``B''. Additionally, the lowest excitation line of HNCO shows
  extended emission suggestive of its presence in a cold extended
  envelope. The relatively higher abundances with respect to CH$_3$OH
  in source ``A'' may originate from higher initial abundances of
  OCN$^-$ in the ice.
\item For CH$_2$CO and C$_2$H$_5$OH only one line is detected due to
  compact emission. For CH$_2$CO these lines are detected with similar
  strength toward both sources and C$_2$H$_5$OH is clearly detected in
  source ``B'', but due to line-blending it is difficult to determine
  the flux for source ``A''. Compact hot emission for CH$_3$CHO is
  detected only toward source ``B''. Comparison with previous single
  dish observations by \citet{cazaux2003} suggests that a cold
  extended component is present as well. If CH$_2$CO, CH$_3$CHO and
  C$_2$H$_5$OH are related through successive hydrogenation on the
  surfaces of grains, the same spatial behavior is expected for all
  three species. Since this is not observed, it suggests that
  hydrogenation reactions on grain surfaces alone cannot account for
  the observed gas phase abundance ratios. The difference between the
  two IRAS~16292-2422 sources can be explained if CH$_3$CHO would be
  selectively destroyed in source ``A'' right before or after
  grain-mantle evaporation.
\end{itemize}
The discussion in this paper demonstrates the strength of clearly
resolved interferometric observations for studies of the chemistry in
star forming regions. First, molecules that have both extended and
compact emission can easily be identified based on $(u,v)$-visibility
analysis. In single dish observations these components will be
averaged together giving rotational temperatures in between that of the
hot and cold component. Additionally, compact emission arising from
different components in the same single-dish beam can be separated,
which leads in some cases to abundance ratios that can vary over an
order of magnitude for different sources as exemplified in
Table~\ref{comp} for HNCO and CH$_3$CHO. For HNCO the presence of a
cold extended component could not be derived from the single dish
observations, whereas for CH$_3$CHO it is the hot component that is
not detected. Further interferometric studies are needed to elucidate
the chemical relations between complex organics.

\begin{acknowledgements}
  We thank Floris van der Tak and Xander Tielens for carefully reading
  this manuscript and an anonymous referee and the editor Malcolm
  Walmsley for constructive comments on the paper. Funding was
  provided by NOVA, the Netherlands Research School for Astronomy and
  a Spinoza grant from the Netherlands Organization for Scientific
  Research (NWO).
\end{acknowledgements}

\Online
\begin{appendix}
\section{SMA Spectra}\label{ap}
\begin{figure}\centering
\includegraphics[width=7.5cm]{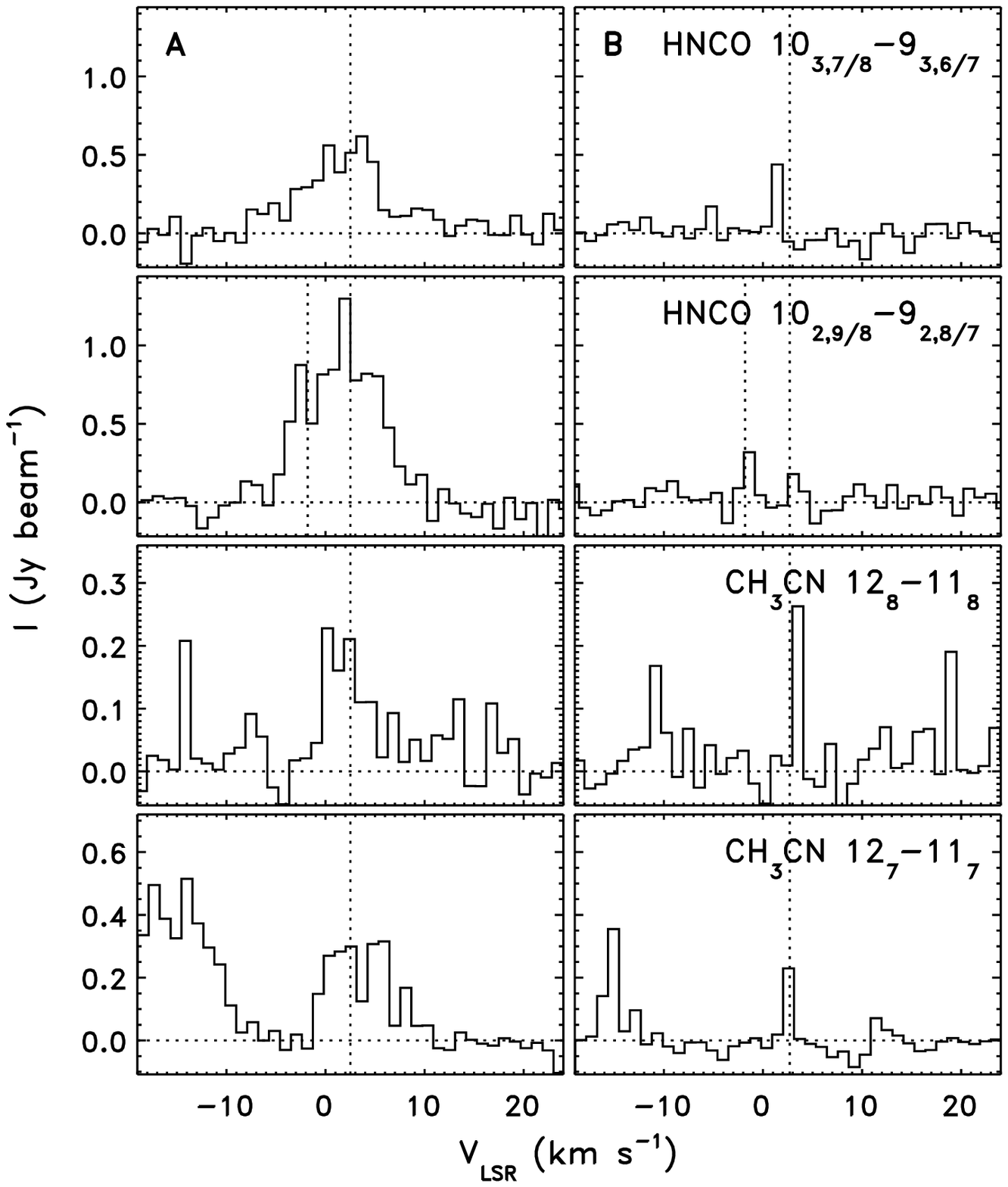}
\caption{Individual SMA spectra for the 10$_{3,7/8}$--9$_{3,6/7}$,
  10$_{2,9}$--9$_{2,8}$ and 10$_{2,8}$--9$_{2,7}$ HNCO transitions and
  the 12$_8$--11$_8$ and 12$_7$--11$_7$ CH$_3$CN transitions. The
  spectra for the ``A'' position are depicted in the left column and
  for ``B'' in the right column. The vertical dotted line indicates
  the average systemic line position. The horizontal dotted line shows
  the baseline.}\label{ap1}
\end{figure}

\begin{figure}\centering
\includegraphics[width=7.5cm]{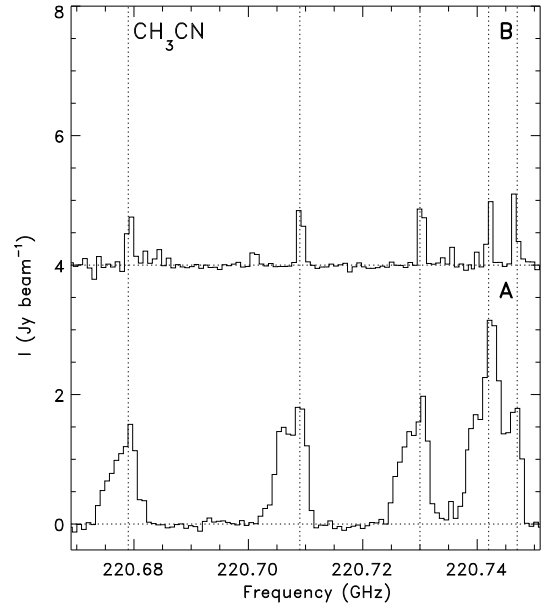}
\caption{SMA spectrum in the 220.668--220.752~GHz range for sources
  ``A'' (lower spectrum) and ``B'' (upper spectrum). The vertical
  dotted lines indicate the positions of, from left to right, the
  12$_4$--11$_4$, 12$_3$--11$_3$, 12$_2$--11$_2$, 12$_1$--11$_1$ and
  12$_0$--11$_0$ CH$_3$CN transitions and the horizontal dotted lines
  the baseline.}\label{ap2}
\end{figure}

\begin{figure}\centering
\includegraphics[width=7.5cm]{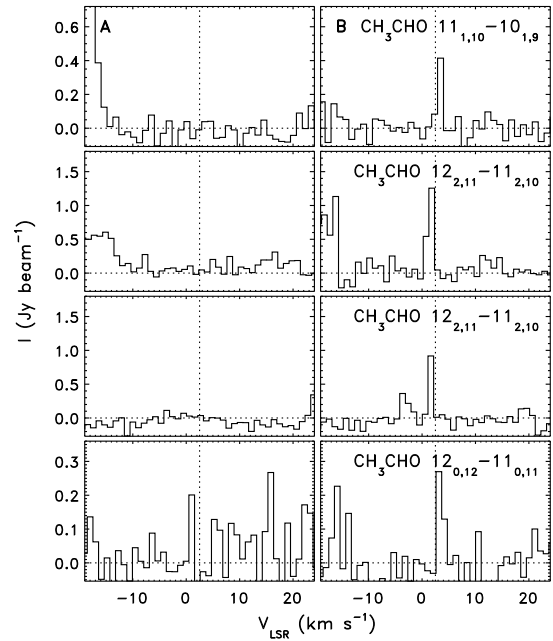}
\caption{Individual SMA spectra for, from left to right, the
  11$_{1,10}$--10$_{1,9}$, 12$_{2,11}$--11$_{2,10}$,
  12$_{2,11}$--11$_{2,10}$, 12$_{0,12}$--11$_{0,11}$ and
  12$_{5,8/8}$--11$_{5,7/6}$ CH$_3$CHO-A transitions.  The spectra for
  the ``A'' position are depicted in the left column and for ``B'' in
  the right column. The horizontal dotted line shows the
  baseline.}\label{ap3}
\end{figure}

\begin{figure}\centering
\includegraphics[width=7.5cm]{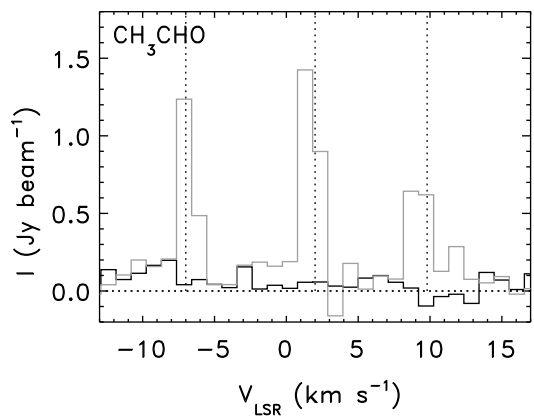}
\caption{SMA spectra for the 12$_{3,9}$--11$_{3,8}$,
  12$_{5,7}$--11$_{5,6}$, and 12$_{5,8}$--11$_{5,7}$ CH$_3$CHO-E
  transitions (dotted lines). The black line indicates the emission
  spectrum toward ``A'' and the grey line toward ``B''. The horizontal
  dotted line shows the baseline.}\label{ap4}
\end{figure}

\end{appendix}

\begin{thebibliography}{52}
\expandafter\ifx\csname natexlab\endcsname\relax\def\natexlab#1{#1}\fi

\bibitem[{{Adams} {et~al.}(1987){Adams}, {Lada}, \& {Shu}}]{adams1987}
{Adams}, F.~C., {Lada}, C.~J., \& {Shu}, F.~H. 1987, \apj, 312, 788

\bibitem[{{Allamandola} {et~al.}(1999){Allamandola}, {Bernstein}, {Sandford},
  \& {Walker}}]{allamandola1999}
{Allamandola}, L.~J., {Bernstein}, M.~P., {Sandford}, S.~A., \& {Walker}, R.~L.
  1999, Space Science Reviews, 90, 219

\bibitem[{{Bis\-schop} {et~al.}(2007{\natexlab{a}}){Bis\-schop}, {Fuchs}, {van
  Dishoeck}, \& {Linnartz}}]{bisschop2007c}
{Bis\-schop}, S.~E., {Fuchs}, G.~W., {van Dishoeck}, E.~F., \& {Linnartz}, H.
  2007{\natexlab{a}}, \aap, 474, 1061

\bibitem[{{Bis\-schop} {et~al.}(2007{\natexlab{b}}){Bis\-schop},
  {J{\o}rgensen}, {van Dishoeck}, \& {de Wachter}}]{bisschop2007a}
{Bis\-schop}, S.~E., {J{\o}rgensen}, J.~K., {van Dishoeck}, E.~F., \& {de
  Wachter}, E.~B.~M. 2007{\natexlab{b}}, \aap, 465, 913

\bibitem[{{Blake} {et~al.}(1994){Blake}, {van Dishoeck}, {Jansen}, {Groesbeck},
  \& {Mundy}}]{blake1994}
{Blake}, G.~A., {van Dishoeck}, E.~F., {Jansen}, D.~J., {Groesbeck}, T.~D., \&
  {Mundy}, L.~G. 1994, \apj, 428, 680

\bibitem[{{Bottinelli} {et~al.}(2007{\natexlab{a}}){Bottinelli}, {Boogert},
  {van Dishoeck}, {Oberg}, {Pontoppidan}, {Blake}, {Evans}, \&
  {Lahuis}}]{bottinelli2007b}
{Bottinelli}, S., {Boogert}, A.~C.~A., {van Dishoeck}, E.~F., {et~al.}
  2007{\natexlab{a}}, in Molecules in Space and Laboratory, ed. J.~L. {Lemaire}
  \& F.~{Combes} ({S. Diana})

\bibitem[{{Bottinelli} {et~al.}(2004{\natexlab{a}}){Bottinelli}, {Ceccarelli},
  {Lefloch}, {Williams}, {Castets}, {Caux}, {Cazaux}, {Maret}, {Parise}, \&
  {Tielens}}]{bottinelli2004b}
{Bottinelli}, S., {Ceccarelli}, C., {Lefloch}, B., {et~al.} 2004{\natexlab{a}},
  \apj, 615, 354

\bibitem[{{Bottinelli} {et~al.}(2004{\natexlab{b}}){Bottinelli}, {Ceccarelli},
  {Neri}, {Williams}, {Caux}, {Cazaux}, {Lefloch}, {Maret}, \&
  {Tielens}}]{bottinelli2004}
{Bottinelli}, S., {Ceccarelli}, C., {Neri}, R., {et~al.} 2004{\natexlab{b}},
  \apjl, 617, L69

\bibitem[{{Bottinelli} {et~al.}(2007{\natexlab{b}}){Bottinelli}, {Ceccarelli},
  {Williams}, \& {Lefloch}}]{bottinelli2007}
{Bottinelli}, S., {Ceccarelli}, C., {Williams}, J.~P., \& {Lefloch}, B.
  2007{\natexlab{b}}, \aap, 463, 601

\bibitem[{{Cazaux} {et~al.}(2003){Cazaux}, {Tielens}, {Ceccarelli}, {Castets},
  {Wakelam}, {Caux}, {Parise}, \& {Teyssier}}]{cazaux2003}
{Cazaux}, S., {Tielens}, A.~G.~G.~M., {Ceccarelli}, C., {et~al.} 2003, \apj,
  593, L51

\bibitem[{{Ceccarelli} {et~al.}(1999){Ceccarelli}, {Caux}, {Loinard},
  {Castets}, {Tielens}, {Molinari}, {Liseau}, {Saraceno}, {Smith}, \&
  {White}}]{ceccarelli1999}
{Ceccarelli}, C., {Caux}, E., {Loinard}, L., {et~al.} 1999, \aap, 342, L21

\bibitem[{{Ceccarelli} {et~al.}(2000){Ceccarelli}, {Loinard}, {Castets},
  {Tielens}, \& {Caux}}]{ceccarelli2000}
{Ceccarelli}, C., {Loinard}, L., {Castets}, A., {Tielens}, A.~G.~G.~M., \&
  {Caux}, E. 2000, \aap, 357, L9

\bibitem[{{Chandler} {et~al.}(2005){Chandler}, {Brogan}, {Shirley}, \&
  {Loinard}}]{chandler2005}
{Chandler}, C.~J., {Brogan}, C.~L., {Shirley}, Y.~L., \& {Loinard}, L. 2005,
  \apj, 632, 371

\bibitem[{{Fuchs} {et~al.}(2008){Fuchs}, {Ioppolo}, {Bisschop}, {Van Dishoeck},
  \& {Linnartz}}]{fuchs2007}
{Fuchs}, G.~W., {Ioppolo}, S., {Bisschop}, S.~E., {Van Dishoeck}, E.~F., \&
  {Linnartz}, H. 2008, submitted to \aap

\bibitem[{{Garrod} {et~al.}(2008){Garrod}, {Widicus Weaver}, \&
  {Herbst}}]{garrod2008}
{Garrod}, R.~T., {Widicus Weaver}, S.~L., \& {Herbst}, E. 2008, accepted by
  \apj, ArXiv e-prints:0803.1214

\bibitem[{{Gibb} {et~al.}(2004){Gibb}, {Whittet}, {Boogert}, \&
  {Tielens}}]{gibb2004}
{Gibb}, E.~L., {Whittet}, D.~C.~B., {Boogert}, A.~C.~A., \& {Tielens},
  A.~G.~G.~M. 2004, \apjs, 151, 35

\bibitem[{{Goldsmith} \& {Langer}(1999)}]{goldsmith1999}
{Goldsmith}, P.~F. \& {Langer}, W.~D. 1999, \apj, 517, 209

\bibitem[{{Huang} {et~al.}(2005){Huang}, {Kuan}, {Charnley}, {Hirano},
  {Takakuwa}, \& {Bourke}}]{huang2005}
{Huang}, H.-C., {Kuan}, Y.-J., {Charnley}, S.~B., {et~al.} 2005, AdSpR, 36, 146

\bibitem[{{Ikeda} {et~al.}(2001){Ikeda}, {Ohishi}, {Nummelin}, {Dickens},
  {Bergman}, {Hjalmarson}, \& {Irvine}}]{ikeda2001}
{Ikeda}, M., {Ohishi}, M., {Nummelin}, A., {et~al.} 2001, \apj, 560, 792

\bibitem[{{Ikeda} {et~al.}(2002){Ikeda}, {Ohishi}, {Nummelin}, {Dickens},
  {Bergman}, {Hjalmarson}, \& {Irvine}}]{ikeda2002}
{Ikeda}, M., {Ohishi}, M., {Nummelin}, A., {et~al.} 2002, \apj, 571, 560

\bibitem[{{J{\o}rgensen} {et~al.}(2007){J{\o}rgensen}, {Bourke}, {Myers}, {Di
  Francesco}, {van Dishoeck}, {Lee}, {Ohashi}, {Sch{\"o}ier}, {Takakuwa},
  {Wilner}, \& {Zhang}}]{jorgensen2007}
{J{\o}rgensen}, J.~K., {Bourke}, T.~L., {Myers}, P.~C., {et~al.} 2007, \apj,
  659, 479

\bibitem[{{J{\o}rgensen} {et~al.}(2005{\natexlab{a}}){J{\o}rgensen}, {Bourke},
  {Myers}, {Sch{\"o}ier}, {van Dishoeck}, \& {Wilner}}]{jorgensen2005b}
{J{\o}rgensen}, J.~K., {Bourke}, T.~L., {Myers}, P.~C., {et~al.}
  2005{\natexlab{a}}, \apj, 632, 973

\bibitem[{{J{\o}rgensen} {et~al.}(2005{\natexlab{b}}){J{\o}rgensen}, {Lahuis},
  {Sch{\"o}ier}, {van Dishoeck}, {Blake}, {Boogert}, {Dullemond}, {Evans},
  {Kessler-Silacci}, \& {Pontoppidan}}]{jorgensen2005a}
{J{\o}rgensen}, J.~K., {Lahuis}, F., {Sch{\"o}ier}, F.~L., {et~al.}
  2005{\natexlab{b}}, \apjl, 631, L77

\bibitem[{{J{\o}rgensen} {et~al.}(2005{\natexlab{c}}){J{\o}rgensen}, {Sch{\"
  o}ier}, \& {van Dishoeck}}]{jorgensen2005}
{J{\o}rgensen}, J.~K., {Sch{\" o}ier}, F.~L., \& {van Dishoeck}, E.~F.
  2005{\natexlab{c}}, \aap, 435, 177

\bibitem[{{J{\o}rgensen} {et~al.}(2002){J{\o}rgensen}, {Sch{\"o}ier}, \& {van
  Dishoeck}}]{jorgensen2002}
{J{\o}rgensen}, J.~K., {Sch{\"o}ier}, F.~L., \& {van Dishoeck}, E.~F. 2002,
  \aap, 389, 908

\bibitem[{{Keane}(2001)}]{keane2001}
{Keane}, J.~V. 2001, PhD thesis, Rijks Universiteit Groningen

\bibitem[{{Kleiner} {et~al.}(1996){Kleiner}, {Lovas}, \&
  {Godefroid}}]{kleiner1996}
{Kleiner}, I., {Lovas}, F.~J., \& {Godefroid}, M. 1996, J. Phys. Chem. Ref.
  Data, 25, 1113

\bibitem[{{Kuan} {et~al.}(2004){Kuan}, {Huang}, {Charnley}, {Hirano},
  {Takakuwa}, {Wilner}, {Liu}, {Ohashi}, {Bourke}, {Qi}, \& {Zhang}}]{kuan2004}
{Kuan}, Y.-J., {Huang}, H.-C., {Charnley}, S.~B., {et~al.} 2004, \apjl, 616,
  L27

\bibitem[{{Liu}(2005)}]{liu2005}
{Liu}, S.-Y. 2005, in IAU Symposium 231, ed. D.~C. {Lis}, G.~A. {Blake}, \&
  E.~{Herbst}, 217--226

\bibitem[{{Mundy} {et~al.}(1992){Mundy}, {Wootten}, {Wilking}, {Blake}, \&
  {Sargent}}]{mundy1992}
{Mundy}, L.~G., {Wootten}, A., {Wilking}, B.~A., {Blake}, G.~A., \& {Sargent},
  A.~I. 1992, \apj, 385, 306

\bibitem[{{Nummelin} {et~al.}(2000){Nummelin}, {Bergman}, {Hjalmarson},
  {Friberg}, {Irvine}, {Millar}, {Ohishi}, \& {Saito}}]{nummelin2000}
{Nummelin}, A., {Bergman}, P., {Hjalmarson}, {\AA}., {et~al.} 2000, \apjs, 128,
  213

\bibitem[{{Olmi} {et~al.}(1993){Olmi}, {Cesaroni}, \& {Walmsley}}]{olmi1993}
{Olmi}, L., {Cesaroni}, R., \& {Walmsley}, C.~M. 1993, \aap, 276, 489

\bibitem[{{Qi}(2006)}]{qi2006}
{Qi}, C. 2006, The MIR Cookbook, The Submillimeter Array/Harvard Smithsonian
  Center for Astrophysicis

\bibitem[{{Remijan} \& {Hollis}(2006)}]{remijan2006}
{Remijan}, A.~J. \& {Hollis}, J.~M. 2006, \apj, 640, 842

\bibitem[{{Sakai} {et~al.}(2006){Sakai}, {Sakai}, \& {Yamamoto}}]{sakai2006}
{Sakai}, N., {Sakai}, T., \& {Yamamoto}, S. 2006, \pasj, 58, L15

\bibitem[{{Sakai} {et~al.}(2007){Sakai}, {Sakai}, \& {Yamamoto}}]{sakai2007}
{Sakai}, N., {Sakai}, T., \& {Yamamoto}, S. 2007, \apj, 660, 363

\bibitem[{{Sch{\"o}ier} {et~al.}(2002){Sch{\"o}ier}, {J{\o}rgensen}, {van
  Dishoeck}, \& {Blake}}]{schoier2002}
{Sch{\"o}ier}, F.~L., {J{\o}rgensen}, J.~K., {van Dishoeck}, E.~F., \& {Blake},
  G.~A. 2002, \aap, 390, 1001

\bibitem[{{Sch{\"o}ier} {et~al.}(2004){Sch{\"o}ier}, {J{\o}rgensen}, {van
  Dishoeck}, \& {Blake}}]{schoier2004}
{Sch{\"o}ier}, F.~L., {J{\o}rgensen}, J.~K., {van Dishoeck}, E.~F., \& {Blake},
  G.~A. 2004, \aap, 418, 185

\bibitem[{{Shirley} {et~al.}(2002){Shirley}, {Evans}, \&
  {Rawlings}}]{shirley2002}
{Shirley}, Y.~L., {Evans}, II, N.~J., \& {Rawlings}, J.~M.~C. 2002, \apj, 575,
  337

\bibitem[{{Sutton} {et~al.}(1985){Sutton}, {Blake}, {Masson}, \&
  {Phillips}}]{sutton1985}
{Sutton}, E.~C., {Blake}, G.~A., {Masson}, C.~R., \& {Phillips}, T.~G. 1985,
  \apjs, 58, 341

\bibitem[{{Takakuwa} {et~al.}(2007){Takakuwa}, {Ohashi}, {Bourke}, {Hirano},
  {Ho}, {J{\o}rgensen}, {Kuan}, {Wilner}, \& {Yeh}}]{takakuwa2007}
{Takakuwa}, S., {Ohashi}, N., {Bourke}, T.~L., {et~al.} 2007, \apj, 662, 431

\bibitem[{{Tielens} \& {Charnley}(1997)}]{tielens1997}
{Tielens}, A.~G.~G.~M. \& {Charnley}, S.~B. 1997, Origins Life Evol. B., 27, 23

\bibitem[{{Turner}(1991)}]{turner1991}
{Turner}, B.~E. 1991, \apjs, 76, 617

\bibitem[{{van Broekhuizen} {et~al.}(2004){van Broekhuizen}, {Keane}, \&
  {Schutte}}]{broekhuizen2004}
{van Broekhuizen}, F.~A., {Keane}, J.~V., \& {Schutte}, W.~A. 2004, \aap, 415,
  425

\bibitem[{{van Broekhuizen} {et~al.}(2005){van Broekhuizen}, {Pontoppidan},
  {Fraser}, \& {van Dishoeck}}]{broekhuizen2005a}
{van Broekhuizen}, F.~A., {Pontoppidan}, K.~M., {Fraser}, H.~J., \& {van
  Dishoeck}, E.~F. 2005, \aap, 441, 249

\bibitem[{{van der Tak} {et~al.}(2003){van der Tak}, {Boonman}, {Braakman}, \&
  {van Dishoeck}}]{vdtak2003}
{van der Tak}, F.~F.~S., {Boonman}, A.~M.~S., {Braakman}, R., \& {van
  Dishoeck}, E.~F. 2003, \aap, 412, 133

\bibitem[{{van der Tak} {et~al.}(2000){van der Tak}, {van Dishoeck}, \&
  {Caselli}}]{vdtak2000b}
{van der Tak}, F.~F.~S., {van Dishoeck}, E.~F., \& {Caselli}, P. 2000, \aap,
  361, 327

\bibitem[{{van Dishoeck} {et~al.}(1995){van Dishoeck}, {Blake}, {Jansen}, \&
  {Groesbeck}}]{dishoeck1995}
{van Dishoeck}, E.~F., {Blake}, G.~A., {Jansen}, D.~J., \& {Groesbeck}, T.~D.
  1995, \apj, 447, 760

\bibitem[{{Watanabe} {et~al.}(2004){Watanabe}, {Nagaoka}, {Shiraki}, \&
  {Kouchi}}]{watanabe2004}
{Watanabe}, N., {Nagaoka}, A., {Shiraki}, T., \& {Kouchi}, A. 2004, \apj, 616,
  638

\bibitem[{{Wilson} \& {Rood}(1994)}]{wilson1994}
{Wilson}, T.~L. \& {Rood}, R. 1994, \araa, 32, 191

\bibitem[{{Yeh} {et~al.}(2008){Yeh}, {Hirano}, {Bourke}, {Ho}, {Lee}, {Ohashi},
  \& {Takakuwa}}]{yeh2008}
{Yeh}, S.~C.~C., {Hirano}, N., {Bourke}, T.~L., {et~al.} 2008, \apj, 675, 454

\bibitem[{{Zinchenko} {et~al.}(2000){Zinchenko}, {Henkel}, \&
  {Mao}}]{zinchenko2000}
{Zinchenko}, I., {Henkel}, C., \& {Mao}, R.~Q. 2000, \aap, 361, 1079

\end{thebibliography}
\end{document}